\documentclass[11pt]{article}
\pdfoutput=1
\usepackage{jheppubmod}
\usepackage[punctsep]{collref}
\usepackage{graphicx}
\usepackage{amssymb}
\usepackage{bbm}
\usepackage{microtype}
\usepackage{mathrsfs}
\usepackage{amsmath,amssymb,mathtools}
\usepackage{slashed}
\usepackage{hyperref}
\usepackage{circuitikz}
\usepackage{caption}
\usepackage{xcolor}
\usepackage{dsfont}
\usepackage{verbatim}
\usepackage{subfig}
\usepackage{amsmath,amssymb,amscd,amsfonts,mathtools,csquotes}
\usepackage{xifthen}
\usepackage{xcolor}
\definecolor{markgreen}{RGB}{230,243,230}
\definecolor{darkolivegreen}{rgb}{0.33, 0.42, 0.18}
\definecolor{darkpastelgreen}{rgb}{0.01, 0.75, 0.24}
\DeclareMathOperator{\Tr}{Tr}

\usepackage[toc,page]{appendix}
\usepackage{epsfig}
\usepackage{epstopdf}
\usepackage{latexsym}
\usepackage{graphicx}
\usepackage{booktabs}
\usepackage{bbm}
\usepackage{color}
\usepackage{physics}
\usepackage{tensor}
\usepackage{verbatim}
\usepackage{subfig}
\usepackage{tikz}
\usepackage{ifthen}
\usepackage{array}
\usetikzlibrary{matrix}
\usetikzlibrary{decorations.markings,calc,shapes,decorations.pathmorphing,patterns,decorations.pathreplacing,arrows,arrows.meta}
\usetikzlibrary{positioning}
\usepackage{hyperref}
\usetikzlibrary{patterns}
\usetikzlibrary{positioning}
\newdimen\mydim
\pdfoutput=1
\makeatletter
\def\@fpheader{\relax}

\newcommand*{\ov}[1]{
  $\m@th\overline{\mbox{#1}}$
}
\newcommand*{\ovA}[1]{
  $\m@th\overline{\mbox{#1}\raisebox{3mm}{}}$
}
\newcommand*{\ovB}[1]{
  $\m@th\overline{\mbox{#1\rule{0pt}{3mm}}}$
}
\newcommand*{\ovC}[1]{
  $\m@th\overline{\mbox{#1\strut}}$
}
\newcommand*{\ovD}[1]{
  $\m@th\overline{\mbox{#1\vphantom{\"A}}}$
}
\newcommand*{\ovE}[1]{
  $\m@th\overline{\raisebox{0pt}[1.2\height]{#1}}$
}
\newcommand*{\ovF}[1]{
  $\m@th\overline{\raisebox{0pt}[\dimexpr\height+1mm\relax]{#1}}$
   Package `calc' can be used as alternative for `\dimexpr'.
}
\newcommand*{\ovG}[1]{
  $\m@th\overline{\raisebox{0pt}[\dimexpr\height+1mm\relax]{#1\vphantom{A}}}$
}
\makeatother
\newboolean{showcomments}
\setboolean{showcomments}{false}
\newcommand\rem[1]{\ifthenelse{\boolean{showcomments}}{{#1}}{}}
\newcommand{\be}{\begin{equation}}
\newcommand{\ee}{\end{equation}}

\usepackage{xifthen}
\newcommand{\dalembert}[1][]{\ifthenelse{\isempty{#1}}{\Box}{#1\Box}}
\usetikzlibrary{decorations.pathmorphing}
\tikzset{snake it/.style={decorate, decoration=snake}}

\newcommand{\E}{{\mathrm e}}

\title{\Large Fermion-Monopole Scattering: A Unified Perspective}
\author{Hao Geng$^{a}$, Zekai Wang$^{b}$}
\affiliation{$^{a}$Gravity, Spacetime, and Particle Physics (GRASP) Initiative, Harvard University, 17 Oxford St., Cambridge, MA, 02138, USA.}
\affiliation{$^{b}$Department of Physics, Harvard University, 17 Oxford St., Cambridge, MA, 02138, USA.}
\emailAdd{haogeng@fas.harvard.edu, zekai$\_$wang@g.harvard.edu}

\abstract{We revisit the scattering of massless chiral fermions by a minimally charged BPS 't Hooft-Polyakov monopole. Existing proposals emphasize different aspects of this process and their connections are often unclear, giving rise to apparent conflicts concerning the outgoing states, energy conservation, and anomalous scattering amplitudes. We clarify the connection between different proposals within a common low-energy fermion-rotor description. Starting from the monopole background, we derive the rotor coupling and retain explicit fermion operators throughout the analysis. By adjoining a decoupled auxiliary scalar, we construct a dual theory that permits exact quantization. We show that the rotor vertex has a continuous spectral support down to zero energy and that its infrared exchange relations identify it as the twist dressing of the outgoing fermion. We also compute anomalous fermion correlators directly from fermion zero modes generated by winding configurations of the compact rotor. These results connect rotor dynamics, twisted outgoing states, and the zero-mode origin of anomalous correlators, resolving apparent conflicts and showing how the existing proposals fit into a mutually consistent picture within their common regime of applicability.} 

\begin{document}
\maketitle
\flushbottom
\pagebreak

\section{Introduction}

The scattering of massless fermions by magnetic monopoles provides a setting in which anomalies, charge conservation, and the structure of asymptotic states are closely intertwined. The Callan–Rubakov effect \cite{Rubakov:1981rg,Rubakov:1982fp,Callan:1982ah,Callan:1982au} illustrates how a monopole can mediate fermion-number-violating processes at energies far below the energy scale required to probe the structure of its core. Even in simple gauge theories, however, the interpretation of the scattering process is subtle. In the lowest angular momentum channel, the allowed incoming and outgoing fermions carry different quantum numbers, e.g. the electric charge. Understanding their relation requires treating the monopole’s electric collective coordinate and the fermionic excitations as a coupled quantum system. The importance of this coupling was already apparent in the early analyses of monopole charge and dyon–fermion dynamics \cite{Callan:1982ah,Callan:1982au}.  See \cite{Yamagishi:1982wp,Affleck:1993np,Maldacena:1995pq,Craigie:1983vq} for some earlier works on the Callan-Rubakov effect.

This problem has recently attracted renewed attention from several perspectives. Descriptions based on generalized symmetries identify outgoing states in twisted sectors, with topological defects dressing accounting for their unconventional quantum numbers \cite{vanBeest:2023dbu,vanBeest:2023mbs}. Analyses of the fermion–rotor model \cite{Polchinski:1984uw} relate this structure to a rotor operator dressing the outgoing fermion \cite{Loladze:2024ayk,Loladze:2025jsq}, or a rotor-dependent varying Fock space of fermions \cite{Hamada:2022eiv}. Other approaches emphasize the dynamics of electric charge in the monopole core \cite{Brennan:2023tae}, study the bound state energy of monopole-fermion system \cite{Hook:2024vhf}, or the role of topologically nontrivial gauge configurations and their fermion zero modes in the anomalous scattering amplitudes \cite{Csaki:2024ajo}. These proposals and the corresponding approaches are further explored and generalized in various works.  See \cite{Csaki:2020inw,Csaki:2020yei,Csaki:2021ozp,Csaki:2022ebw,Csaki:2022tvb,Khoze:2023kiu} for pairwise amplitude based approaches, \cite{Bogojevic:2024kch,Bogojevic:2024xtx,Koundouros:2026rnb}, for monopole/dyon and fermion--rotor dynamics, \cite{Hamada:2026jgj} for the construction of semiton states, and \cite{Tachikawa:2026cxd,Arias-Tamargo:2026urw,Wei:2026fsn,Arias-Tamargo:2026kbv,Ohmori:2026tip,Cordova:2026dtw,Chen:2023hmm,Wei:2025zyd,Copetti:2024rqj,Copetti:2024dcz,Shimamori:2025ntq,Ueda:2025ecm,Antinucci:2026uuh,Garner:2023izn,Garner:2024tis} for boundary, defect, and twistorial approaches to this type of scattering processes.
For broader context, see \cite{Wiesiolek:2022oeb,Chen:2023czk,Bah:2025zzk,Bartsch:2026wqq,Cordova:2023ent,Sun:2023xxv,Cordova:2023bja,Antinucci:2023ezl,Nagoya:2023zky,Choi:2023vgk,Nagoya:2024gtq,Kan:2024fuu,Brennan:2023kpw,DelZotto:2024tae,Nardoni:2024sos,Copetti:2024onh,Copetti:2026ncv,Harder:2026csj,Bhardwaj:2023idu,Bhardwaj:2023bbf,Bhardwaj:2024qrf,Bhardwaj:2024ydc,Wen:2025thg,Lu:2025rwd} for generalized symmetries and their phases, \cite{Khoze:2024hlb,Bolognesi:2024kkb,Candela:2025gwp,Craig:2026rdf,Cordova:2023her,Cordova:2024ypu,DelZotto:2024ngj,Delgado:2024pcv,Platschorre:2025byo,Choi:2025vxr,Hong:2025qbw,Choi:2026oqz,Li:2024nuo,Koren:2024xof,Delgado:2026lwj,Cordova:2022ieu,Heidenreich:2023pbi} for particle-physics applications, \cite{Cvetic:2023pgm,Gould:2024yzt,Gutperle:2024vyp,Bharadwaj:2024gpj,Zeng:2024tuq,Xu:2025hfs,Zakharov:2026nly,Lu:2026itw,Sarma:2026bqi,Davighi:2023mzg,Berean-Dutcher:2025ohp,Derda:2025flq,Tizzano:2026rgr,Aharony:2023amq,Kristjansen:2024map,Freivogel:2026ujn,ihojin:2025pjj,Guo:2026mcc} for developments in other field-theoretic settings, \cite{Bhardwaj:2023kri,Shao:2023gho,Galati:2023yxy,Choi:2024bcq,Rizi:2024gsj,Schafer-Nameki:2024wvw,McCullough:2024evr,RiedelGarding:2025vqd,Kaidi:2026urc,Onder:2024vmb} for reviews and broader expositions.

However, the connection between these proposals is not manifest. Their accounts of the outgoing state, the transfer of electric charge, and the origin of anomalous processes can appear to conflict. It is therefore important to determine which differences reflect distinct physical assumptions and which arise from different descriptions of the same dynamics.

The central aim of this paper is to clarify the connection and resolve the apparent conflicts between the above proposals. We develop a dual description in which the system is exactly solvable. Within the low-energy regime where the above proposals overlap, we show how their principal mechanisms fit into a mutually consistent account of the scattering process. Making these relations explicit also identifies the assumptions and operator definitions needed to compare the different formulations.

We revisit the minimal setting of fermion-monopole scattering by considering a minimally charged BPS ’t Hooft–Polyakov monopole coupled to an even number ($N_f=2N$) of massless left-handed fermions in the fundamental representation of the gauge group $SU(2)$. We work in the weak-coupling regime of this gauge theory and focus on energies far below the inverse monopole core size. Starting from the monopole background, we identify the electric collective coordinate as the phase of the charged gauge field condensation in the core and derive its coupling to the lowest angular momentum fermions. Unfolding the radial problem gives a 2d chiral fermion theory coupled to a compact rotor degree of freedom. Its contact limit provides the tractable fermion–rotor model proposed in \cite{Polchinski:1984uw}. 

Staring from the fermion-rotor model, we work out the fermion equation of motion which makes the role of the rotor particularly transparent. Propagation through the core relates an incoming fermion to an outgoing fermion dressed by a rotor vertex evaluated at the time of passage through the monopole \cite{Loladze:2024ayk,Loladze:2025jsq}. This dressing accounts for the transfer of electric charge. It also raises an apparent energy puzzle: an isolated rotor with a moment of inertia $I$ has an energetic excitation gap of order $I^{-1}$, whereas the scattering process can occur at arbitrarily low energy of the fermion. A comparison of the proposed scattering pictures must therefore account for the spectrum and operator matrix elements of the coupled system.

To address this energy puzzle, we map the system to a new dual description by first adding a decoupled auxiliary scalar to the fermion-rotor Lagrangian. Then, we work out the explicit operator dictionary between the original theory and a dual description of the theory. The dual theory consists of decoupled free fermions, the rotor and another scalar on the half-line whose boundary value is coupled to the rotor dynamics. In this dual description, the bosonic rotor-scalar sector is quadratic and thus exactly solvable. The operator dictionary is complicated and highly non-local, for example, the fermions in the original description are mapped into fermions in the dual description dressed with the bosonic sector at every momentum.

Formulating the fermion-rotor model in the dual description enables us to quantize the system exactly and extract the physics in the original description from this exact solution. We correct a common misconception in the literature that the energy cost for exciting the rotor is of order $I^{-1}$; in fact, we show that it can be arbitrarily small. We demonstrate this result by first showing that quantizing the coupled scalar–rotor sector gives a gapless energy spectrum. Then we compute the spectral representation of the rotor vertex $\E^{i\hat{\alpha}(t)}$ and show that it has support at arbitrarily small excitation energy. The operator dictionary also allows us to calculate the commutation relation between the retarded rotor vertex and the fermion field. In the infrared limit, this relation takes the form of a twist operator action on the outgoing fermions, confirming the results of \cite{vanBeest:2023dbu,Loladze:2025jsq}. Therefore, the dynamical rotor dressing and the twisted-sector interpretation of the outgoing excitation are thereby connected through an explicit operator calculation.

After addressing the energy puzzle, we go back to the original description and compute anomalous fermion correlators in the Euclidean signature with the time compactified. The compactness of the rotor coordinate requires the Euclidean path integral to include configurations with nonzero winding. Such configurations will source fermion zero modes. The number of these zero modes is related to the ABJ anomaly by index theorem, and so fermion insertions saturating these zero modes yield anomalous correlation functions. We evaluate the two-point and four-point anomalous correlation functions as explicit examples, and provide the results for the generalization to higher-point functions. We reproduce essentially the same physics of ``Abelian Instanton" in \cite{Csaki:2024ajo}. In the fermion-rotor language, the anomalous fermion number violating correlators are mediated by rotor winding configuration. This calculation is technical as it requires keeping track of the zero-mode Gram matrix, the reduced fermion determinant, and the path-integral measure. This gives a direct winding-sector calculation of correlators that can also be accessed through cluster decomposition in the fermion–rotor model \cite{Polchinski:1984uw,Loladze:2025jsq}, and makes their topological origin explicit. These correlators are not suppressed by any parameters, and they are the essence of the Callan-Rubakov effect.

Finally, as in \cite{Brennan:2023tae} we show that a field redefinition of fermion operators can redistribute the energy and charge between the bulk fermion field and the rotor, clarifying why equivalent descriptions can suggest different interpretations of charge and energy transfer. Our calculations explain how several seemingly different resolutions of the puzzles in fermion–monopole scattering can coexist. The outgoing fermion twist dressing, the absence of an isolated-rotor energy threshold, and the winding-sector contributions to anomalous correlators emerge within the same effective theory. Our comparison thus establishes concrete connections between different proposals and resolves apparent conflicts in their common domain of applicability. The assumptions concerning dynamical gauge fields, the monopole core, and global charge sectors remain explicit throughout this comparison.

The paper is organized as follows. Sec.~\ref{sec:background} revisits the standard results of monopole background, its electric collective coordinate, the fermion coupling, angular decomposition, and s-wave reduction. For convenience, we introduce two frames which we call the tilded and untilded frames respectively. We set up the scattering problem in both tilded and untilded frames at the end. Sec.~\ref{sec:untildedframe} works in the untilded frame and contains the main novel results of this paper. Starting with the fermion–rotor model, we develop the dual description and its exact quantization. Then we find a continuous rotor energy spectrum and reproduce the result of the twist operator proposal. Then we calculate anomalous fermion correlators in the original description. Sec.~\ref{sec:thetachanging} analyzes the system in the tilded frame. Sec.~\ref{sec:comments} discusses various existing proposals to this problem in the recent literature. We try to build connections and clarify apparent tensions between these proposals. Finally, we conclude in Sec.~\ref{sec:conclusions}. Appendix.~\ref{sec:derivation} derives the dual description and the operator dictionary between the dual description and the original description. Appendix.~\ref{sec:normalization} analyzes the reduced fermion determinants and the fermion zero-mode normalization.

\section{Monopoles, Massless Fermions and Their Interaction}\label{sec:background}
In this section, we set up the conventions and notations in this paper and revisit the fermion-monopole system. We only consider the minimally charged $SU(2)$ 't Hooft-Polyakov monopole \cite{tHooft:1974kcl,Polyakov:1974ek} in the BPS limit coupled to an even number of massless Weyl fermions in the fundamental representation.

\subsection{Notations and Conventions}
Let's first define our notations and conventions. We use the mostly positive signature for the metric in 4d
\begin{equation}
    \eta^{\mu\nu}=\text{diag}(-1,1,1,1)\,.
\end{equation}
We take the generators of the gauge symmetry to be hermitian. The gauge field strength is given in terms of the gauge field as
\begin{equation}
    F_{\mu\nu}^{a}=\partial_{\mu}A^{a}_{\nu}-\partial_{\nu}A^{a}_{\mu}-e\epsilon^{abc}A_{\mu}^{b}A_{\nu}^{c}\,.
\end{equation}
On the adjoint scalar, we have the covariant derivative
\begin{equation}
    D_{i}\Phi^{a}=\partial_{i}\Phi^{a}-e\epsilon^{abc}A_{i}^{b}\Phi^{c}\,.
\end{equation}
The SU(2) gauge generators are
\begin{equation}
    T^{a}=\frac{\sigma^{a}}{2}\,,\quad\text{for $a=1,2,3$}\,,\quad T^{\pm}=\frac{T^{1}\pm i T^{2}}{\sqrt{2}}\,.
\end{equation}
The Dirac $\gamma$-matrices are
\begin{equation}
    \gamma^{0}=\begin{pmatrix}
0 & \mathbf{1}_{2\times2} \\
-\mathbf{1}_{2\times2} & 0
\end{pmatrix}\,,\quad \gamma^{j}=\begin{pmatrix}
0 & \sigma^{j} \\
\sigma^{j} & 0
\end{pmatrix}\,,\quad\gamma^{5}=\begin{pmatrix}
 -\mathbf{1}_{2\times2} &0 \\
0& \mathbf{1}_{2\times2}
\end{pmatrix}\,.
\end{equation}
On a fundamental Dirac fermion, we have the covariant derivative
\begin{equation}
    D_{\mu}=\partial_{\mu}+ieA_{\mu}^{a}T^{a}\,,
\end{equation}
and the conjugate fermion operator is
\begin{equation}
    \bar{\Psi}=-\Psi^{\dagger}\gamma^{0}
\end{equation}
We will consider $N_{f}=2N$ massless Weyl fermions and for each flavor $f$. These fermions are all in the fundamental representation of the $SU(2)$. Thus, when the $SU(2)$ is spontaneously broken to $U(1)$ the charges of the fermions are all $q_{f}=\frac{1}{2}$.  Thus, we will use an abstract $q_{f}$ and $q_{f}=\frac{1}{2}$ interchangeably in this draft.

\subsection{'t Hooft-Polyakov Monopole as Local W-boson Condensation}
In this subsection, we revisit the standard story of the BPS 't Hooft-Polyakov monopole \cite{tHooft:1974kcl,Polyakov:1974ek,tHooft:1999cgx}, with the goal of emphasizing it as a configuration of localized W-boson condensation in the infrared limit. 

We have the SU(2) gauge field $A_{\mu}^{a}$ coupled with an adjoint scalar $\Phi^{a}$ with a potential $V(\Phi)$, along with the following action
\begin{equation}
    \mathcal{L}=-\frac{1}{4} F_{\mu\nu}^{a}F^{a\mu\nu}-\frac{1}{2}D_{\mu}\Phi^{a}D^{\mu} \Phi^{a}-V(\Phi)\,,
\end{equation}
and the scalar potential is
\begin{equation}
    V(\Phi)=\frac{\lambda}{4}(\Phi^{a}\Phi^{a}-v^2)^2\,.
\end{equation}
We consider the limiting case $\lambda\rightarrow0$, such that for finite energy configurations, we have
\begin{equation}
    \Phi^{a}\Phi^{a}\rightarrow v^2,\quad\text{as $r\rightarrow\infty$}\,,
\end{equation}
and in the bulk we can safely take $\lambda=0$. We will take the gauge that $A_{0}^{a}=0$ and the 't Hooft-Polyakov monopole is given by the static maximal symmetric configuration that preserves the diagonal part of the SO(3) gauge symmetry\footnote{On the adjoint representation $SU(2)\simeq SO(3)$.} and the SO(3) rotation 
\begin{equation}
    \Phi^{a}=\frac{\hat{r}^{a}}{er}H(evr)\,\quad A_{i}^{a}=-\frac{\epsilon^{a}_{ij}\hat{r}^{j}}{er}(1-K(evr))\,,\label{eq:ansarz}
\end{equation}
where we have the W-boson mass $m_{W}=ev$ and the boundary conditions
\begin{equation}
    H(evr)\rightarrow evr\,,\quad K(evr)\rightarrow 0\,\quad\text{as $r\rightarrow\infty$}\,,\label{eq:bc}
\end{equation}
This configuration should satisfy the equations of motion
\begin{equation}
    D_{\mu}F^{a\mu\nu}+e\epsilon^{abc}\Phi^{b}D^{\nu}\Phi^{c}=0\,,\quad D_{\mu}D^{\mu}\Phi^{a}=0\,.\label{eq:fulleom}
\end{equation}
In fact, in the above $A_{0}^{a}=0$ gauge, the static property reduced the equations of motion to
\begin{equation}
    D_{i}F^{aij}+e\epsilon^{abc}\Phi^{b}D^{j}\Phi^{c}=0\,\quad D_{i}D^{i}\Phi^{a}=0\,.\label{eq:eom}
\end{equation}
The BPS configuration minimizes the energy
\begin{equation}
    M=\frac{1}{2}\int d^{3}x\Big(B^{ai}B^{a}_{i}+D_{i}\Phi^{a}D^{i}\Phi^{a}\Big)\,,
\end{equation}
which thus obeys
\begin{equation}
    D_{i}\Phi^{a}=B_{i}^{a}=\frac{1}{2}\epsilon_{ijk}F^{a}_{jk}\,.\label{eq:BPS}
\end{equation}
It is easy to see that Equ.~(\ref{eq:BPS}) implies Equ.~(\ref{eq:eom}) using Bianchi identity and \begin{equation}
    [D_{i},D_{j}]\Phi^{a}=-e\epsilon^{abc}F_{ij}^{b}\Phi^{c}\,.
\end{equation}
Thus, we only have to consider Equ.~(\ref{eq:BPS}) and it implies that 
\begin{equation}
    evrK'(evr)=-HK\,,\quad evrH'(evr)=H-(K-1)(K+1)\,.
\end{equation}
As a result, together with the asymptotic boundary conditions Equ.~(\ref{eq:bc}), we have the exact solution
\begin{equation}
    H(m_{W}r)=m_{W}r\coth(m_{W}r)-1\,,\quad K(m_{W}r)=\frac{m_{W}r}{\sinh(m_{W}r)}\,.
\end{equation}

For the convenience of the discussion of the IR limit and the matching of it to QED, let's go to the unitary gauge. This gauge transform on Equ.~(\ref{eq:ansarz}) rotates the adjoint scalar to the third direction in the internal SU(2) directions. This direction is then the unbroken U(1) of the effective QED. The gauge transformation is simply
\begin{equation}
    U(\theta,\phi)=e^{i\theta T^2}e^{i\phi T^{3}}\,,\label{eq:gaugetransform}
\end{equation}
which rotates the radial vector $\vec{r}$ to the $z$-direction. Thus, we have the background configuration in the unitary gauge
\begin{equation}
    \Phi^{a}=\frac{\delta^{a3}}{er}H(evr)\,,\quad A=\frac{K(evr)}{e}\Big(T^{1}\sin\theta d\phi-T^{2}d\theta\Big)-\frac{\cos\theta}{e}T^{3}d\phi\,.
\end{equation}
Written in terms of the photon $A^{3}$ and the W-boson $W^{\pm}$ we have
\begin{equation}
    \bar{A}^{3}=-\frac{\cos\theta}{e}d\phi\,,\quad \bar{W}^{\pm}=\frac{m_{W}r}{\sqrt{2}e\sinh m_{W}r}(\sin\theta d\phi\pm id\theta)\,,\label{eq:background}
\end{equation}
where we have introduced the $\bar{}$ to emphasize that the BPS monopole is a background configuration.
As a result, for $r\gg\frac{1}{m_{W}}$ we have a Dirac monopole from the photon sector and an exponentially vanishing $W$-boson profile. However, for $r\ll\frac{1}{m_{W}}$, the W-bosons are condensed to
\begin{equation}
    \bar{W}^{\pm}\sim\frac{1}{\sqrt{2}e}\Big(\sin\theta d\phi\pm id\theta\Big)\,.\label{eq:Wlocal}
\end{equation}
Hence, the monopole is really a localized W-boson condensation in the IR limit, where $\frac{1}{m_{W}}\rightarrow0$ compared to all other length scales in the theory.

\subsection{The Rotor as the Localized Goldstone Boson}\label{sec:condensation}
Having understood that the W-boson is locally condensed near the monopole core, let's try to work out the low energy effective theory of the monopole. For the sake of convenience to turn on the matter coupling in the next section, let's take the gauge 
\begin{equation}
    W_{0}^{\pm}=0\,,\label{eq:gaugeW}
\end{equation}
but treat the photon $A_{\mu}^{3}$ fully covariantly. Let's treat the monopole position as fixed. The low energy theory is captured by the Goldstone boson from the W-boson condensation in Equ.~(\ref{eq:background}) and dynamical photon. The full Lagrangian in the gauge sector is
\begin{equation}
    \mathcal{L}=-\frac{1}{4}F_{\mu\nu}^{a}F^{a\mu\nu}=-\frac{1}{4}F^{3}_{\mu\nu}F^{3\mu\nu}-\frac{1}{2}F_{\mu\nu}^{+}F^{-\mu\nu}\,.
\end{equation}
Expanding it in terms of the gauge fields we have
\begin{equation}
\begin{split}
    \mathcal{L}=&-\frac{1}{4}F_{\mu\nu}F^{\mu\nu}-\frac{1}{2}W_{\mu\nu}^{+}W^{-\mu\nu}-ieF_{\mu\nu}W^{+\mu}W^{-\nu}\\&+\frac{e^2}{4}\Big(W_{\mu}^{+}W_{\nu}^{-}-W_{\mu}^{-}W_{\nu}^{+}\Big)\Big(W^{+\mu}W^{-\nu}-W^{-\mu}W^{+\nu}\Big)\,,
    \end{split}
\end{equation}
where we have denoted the photon field as $A_{\mu}$ and its Maxwell field strength as $F_{\mu\nu}$ and we have
\begin{equation}
    W_{\mu\nu}^{\pm}=D_{\mu}W^{\pm}_{\nu}-D_{\nu}W^{\pm}_{\mu}\,,\quad D_{\mu}W^{\pm}_{\nu}=\partial_{\mu}W^{\pm}_{\nu}\pm ieA_{\mu}W_{\nu}^{\pm}\,.
\end{equation}
In the low energy limit, we can parametrize the field as 
\begin{equation}
    W_{\mu}^{\pm}=e^{\pm i\theta(x,t)}\bar{W}_{\mu}^{\pm}\,,\quad A_{\mu}=\bar{A}_{\mu}+a_{\mu}\,.\label{eq:Wfield}
\end{equation}
Since we are consistently in the gauge Equ.~(\ref{eq:gaugeW}) for both the background Equ.~(\ref{eq:background}) and the low energy fluctuation Equ.~(\ref{eq:Wfield}), we have to impose the Gauss' law constraints by hands. The Gauss' law constraints associated with both the W-boson and the photon are:
\begin{equation}
\begin{split}
    &D_{i}F^{0i\pm}=\partial_{i}F^{\pm0i}\pm ieA_{i}F^{\pm 0i}\mp ieW_{i}^{\pm}F^{0i3}=0\,,\\
   &D_{i}F^{0i3}=\partial_{i}F^{0i3}+ieW_{i}^{+}F^{0i-}-ieW_{i}^{-}F^{0i+}=0\,.\label{eq:constraints}
    \end{split}
\end{equation}
We can impose the above constraints in the background Equ.~(\ref{eq:background}) for the field Equ.~(\ref{eq:Wfield}). As a side comment, we note that in the full quantum theory, with matter coupled, the constraints also involve the matter part. But for now we want to work out the low energy modes in the monopole sector that obey the constraints by themselves. Then the other modes will enter the constraints with the matter involved, with the result that matter operators should be dressed by these other modes to ensure the gauge invariance of the observables. Though in this paper, we are mostly interested in the $e\rightarrow0$ limit.

Expanding the constraints Equ.~(\ref{eq:constraints}) in the field Equ.~(\ref{eq:Wfield}) we have
\begin{equation}
    \begin{split}
       \pm i\Big(\partial^{0}\partial_{i}\theta+e\partial_{i}a^{0}\Big)\bar{W}^{\pm i}\pm i&\Big(\partial^{0}\theta+ea^{0}\Big)\Big(\partial_{i}\pm i \partial_{i}\theta\pm ie\bar{A}_{i}\pm iea_{i}\Big)\bar{W}^{\pm i}\\&\mp ie\Big(\partial^{0}a^{i}-\partial^{i}a^{0}\Big)\bar{W}^{\pm i}=0\,,\\
       \partial_{i}(\partial^{0}a^{i}-\partial^{i}a_{0})+&2e(\partial^{0}\theta+ea^{0})\bar{W}^{+}_{i}\bar{W}^{-i}=0\,.
    \end{split}
\end{equation}
Using the BPS solution for the background fields, it is easy to see that we have the following exact solution mode
\begin{equation}
    \theta(x,t)=\hat{\alpha}(t)\frac{H(evr)}{ver}\,,\quad a_{0}(x,t)=\hat{a}_{0}(t)\frac{H(evr)}{ver}\,,\quad a_{r}(x,t)=-\hat{\alpha}(t)\partial_{r}\frac{H(evr)}{ve^2 r}\,,\label{eq:solmode}
\end{equation}
where we use $\hat{}$ to denote the operator part. Now we are ready to evaluate the effective Lagrangian of these modes
\begin{equation}
\begin{split}
    \mathcal{L}_{\text{monopole}}&=-\frac{1}{4}(\partial_{\mu}a_{\nu}-\partial_{\nu}a_{\mu})(\partial^{\mu}a^{\nu}-\partial^{\nu}a^{\mu})-(\partial_{\mu}\theta+ea_{\mu})(\partial^{\mu}\theta+ea^{\mu})\bar{W}_{\nu}^{+}\bar{W}^{-\nu}\,,\\&=(\partial_{t}\hat{\alpha}(t)+e\hat{a}_{0}(t))^2\Big(\frac{1}{2}\partial_{i}\frac{H(evr)}{ve^2 r}\partial^{i}\frac{H(evr)}{ve^2 r}+\frac{v^2H^2(evr)}{v^2e^2 r^2\sinh^2(evr)}\Big)\,.
    \end{split}
\end{equation}
Thus, we have the effective action
\begin{equation}
    S_{\text{monopole, naive}}=\frac{2\pi v^2}{m_W^{3}}\int dt (\partial_{t}\hat{\alpha}(t)+e\hat{a}_{0}(t))^2\,.\label{eq:stuckelberg}
\end{equation}
We note that this action is in the St\"{u}ckelberg form if both $\hat{\alpha}(t)$ and the remnant gauge mode $\hat{a}_{0}(t)$ are dynamical. This confirms our intuition that the rotor mode $\hat{\alpha}(t)$ is a Goldstone boson due to the W-boson condensation. However, as we will see in the next subsection, the remnant gauge mode $\hat{a}_{0}(t)$ is not dynamical.
 
\subsection{A Puzzle for the Remnant Gauge Field and Its Resolution}
We have obtained the effective action Equ.~(\ref{eq:stuckelberg}), which however has the following gauge symmetry if both the rotor and the remnant gauge fields are dynamical
\begin{equation}
    \hat{\alpha}(t)\rightarrow\hat{\alpha}(t)+e\delta(t)\,,\quad\hat{a}_{0}(t)\rightarrow\hat{a}_{0}(t)-\dot{\delta}(t)\,.
\end{equation}
This implies that the system is purely topological. This can been seen by computing the finite temperature partition function of this system using path integral
\begin{equation}
    Z(\beta)=\tr e^{-\beta H}=\int\frac{1}{\text{Vol}(\text{gauge})}D\alpha Da^{E}_{0}e^{-\frac{2\pi v^2}{m_{W}^3}\int_{0}^{\beta} d\tau (\partial_{\tau}\hat{\alpha}(\tau)+e\hat{a}^{E}_{0}(\tau))^2}\,,
\end{equation}
where we have used the Euclidean gauge field $a_{0}^{E}(\tau)=-ia_{0}(-i\tau)$. This path integral could be computed using the Faddeev-Popov trick. But here it can be computed more straightforwardly, as we can first integrate over the gauge field and then compute the rotor integral. This gives
\begin{equation}
    Z(\beta)=\text{const.}\times\int \frac{D\alpha}{\text{Vol}(\text{gauge})}=\text{const.}\times1\,,
\end{equation}
which is independent of the inverse temperature $\beta$. Thus, if both of the rotor and the remnant gauge field are dynamical, the resulting effective quantum mechanical system is purely topological with no excitations. This is a bit puzzling if one reminds himself/herself about the Julia-Zee dyon \cite{Julia:1975ff} and its BPS limit. Thus, the monopole should have a tower of electrically charged excitations above it. This tower should be sourced by the rotor $\hat{\alpha}(t)$.

The above puzzle can be resolved if we look more closely at our exact solution mode in Equ.~(\ref{eq:solmode}) and remind ourselves that $\theta(x,t)$ is the phase of the localized W-boson condensation in Equ.~(\ref{eq:Wlocal}). The profile function $\frac{{H(evr)}}{ver}$ approaches to the constant $1$ when $r\rightarrow \infty$. Thus, the remnant gauge mode $\hat{a}_{0}(t)$ is really an asymptotic gauge field and it should not be dynamical. Nevertheless, this doesn't obstruct the rotor mode $\hat{\alpha}(t)$ from being dynamical as $\theta(x,t)$ is the phase of the W-boson and the W-boson is only locally condensed at the monopole core in the IR limit. Therefore, the rotor $\hat{\alpha}(t)$ is not an asymptotic field but the remnant gauge field $\hat{a}_{0}(t)$ is. As a result, the remnant gauge field $\hat{a}_{0}(t)$ should be frozen to zero and the true effective action is
\begin{equation}
    S_{\text{monopole}}=\frac{2\pi v^2}{m_W^{3}}\int dt (\partial_{t}\hat{\alpha}(t))^2\,.
\end{equation}

From Equ.~(\ref{eq:solmode}), the rotor is the zero mode of the Goldstone boson $\theta(x,t)$, i.e. $\theta(x,t)\rightarrow \hat{\alpha}(t)$ as $r\rightarrow\infty$. The Goldstone boson, as the phase of the W-boson, is $2\pi$-periodic. Thus, in the IR limit where $f(r)\rightarrow1$ the rotor is periodically identified with itself as
\begin{equation}
    \hat{\alpha}(t)=\hat{\alpha}(t)+2\pi\,.
\end{equation}

In summary, we have identified the rotor mode $\hat{\alpha}(t)$ as the zero mode of the Goldstone boson due to the local W-boson condensation, while the remnant gauge field $\hat{a}_{0}(t)$ is an asymptotic field which should thus be frozen. This fact helps us to resolve an apparent puzzle.

\subsection{The Fermion-Monopole Hamiltonian}
Now let's turn on the fermion fields. As in the standard model, we want the fermions to be chirally coupled to the SU(2) gauge field in the fundamental representation. To avoid the Witten anomaly \cite{Witten:1982fp}, we will have an even number of massless left-handed fermions $\psi_{f}$ with $f=1,2,\cdots,2N$. This theory thus has an SU(2$N$) flavor symmetry. The Lagrangian of these fermions is
\begin{equation}
    \mathcal{L}_{\text{fermion}}=\bar{\psi}_{f}i\gamma^{\mu}\frac{1-\gamma^5}{2}D_{\mu}\psi_{f}\,,\label{eq:DiracLag}
\end{equation}
and the Hamiltonian is
\begin{equation}
\begin{split}
\mathcal{H}_{\text{fermion}}&=\psi^{\dagger}_{f}(i\sigma^{i}\partial_{i}-e\sigma^{i}A^{a}_{i}T^{a}+eA_{0}^{a}T^{a})\psi_{f}\,,
\end{split}
\end{equation}
for which in terms of the W-boson and the photon we have
\begin{equation}
    A_{\mu}^{a}T^{a}=A_{\mu}T^{3}+W_{\mu}^{+}T_{+}+W_{\mu}^{-}T_{-}\,.
\end{equation}
Thus, in the monopole background Equ.~(\ref{eq:background}), we have
\begin{equation}
    A_{i}^{a}T^{a}=(\bar{A}_{i}+a_{i})T^{3}+e^{i\hat{\alpha}(t)f(r)}\bar{W}_{i}^{+}T_{+}+e^{-i\hat{\alpha}(t)f(r)}\bar{W}_{i}^{-}T_{-}\,,\quad A_{0}^{a}T^{a}=0\,,\quad f(r)\equiv\frac{H(evr)}{ver}\,.
\end{equation}
We want to study the time evolution of the fermion under the above Hamiltonian, which describes the scattering of the fermion by the monopole. We can first take IR limit $m_{W}=ev\rightarrow\infty$ which gives the localized W-boson condensation Equ.~(\ref{eq:Wlocal}). For convenience, let's define $g(r)=\frac{m_{W}r}{\sinh m_{W}r}$, 
then we have
\begin{equation}
    \bar{W}^{\pm}=g(r)\frac{1}{\sqrt{2}e}\Big(\sin\theta d\phi\pm id\theta\Big)\,.
\end{equation}
The final Hamiltonian can be written using the free Hamiltonian and the fermion current operators $J^{a}_{\mu}=\bar{\psi}_{f}\gamma_{\mu}\frac{1-\gamma^5}{2}T^{a}\psi_{f}$ as
\begin{equation}
    \mathcal{H}_{\text{fermion}}=\mathcal{H}_{0}+e (\bar{A}_{i}-\frac{1}{e}\hat{\alpha}(t)\partial_{i}f(r))J^{i}+ ee^{i\hat{\alpha}(t)f(r)}\bar{W}^{+}_{i}J_{+}^{i}+ee^{-i\hat{\alpha}(t)f(r)}\bar{W}_{i}^{-}J_{-}^{i}\,,\label{eq:Hfull}
\end{equation}
Expanding in the spinor space, we have
\begin{equation}
\begin{split}
    \bar{W}_{i}^{\pm}\sigma^{i}&=\pm\frac{ig(r)}{\sqrt{2}er}\begin{pmatrix}
-\sin\theta& (\cos\theta\mp1)e^{-i\phi} \\
(\cos\theta\pm1)e^{i\phi} & \sin\theta
\end{pmatrix}\\&=\pm\frac{ig(r)}{\sqrt{2}er}\Bigg(-\sin\theta\sigma^{3}+\begin{pmatrix}
0& (\cos\theta\mp1)e^{-i\phi} \\
(\cos\theta\pm1)e^{i\phi} & 0\label{eq:Wsing}
\end{pmatrix}\Bigg)\,,
\end{split}
\end{equation}
and
\begin{equation}
    \quad \bar{A}_{i}\sigma^{i}=\frac{i\cos\theta}{er\sin\theta}\begin{pmatrix}
0& e^{-i\phi} \\
-e^{i\phi} & 0
\end{pmatrix}\,.
\end{equation}
We can see that there are singular terms for $r\rightarrow0$ due to the factor $\frac{g(r)}{r}$ in Equ.~(\ref{eq:Wsing}). Thus, we gauge transform back to the regular gauge Equ.~(\ref{eq:ansarz}) using Equ.~(\ref{eq:gaugetransform}), which results in
\begin{equation}
\begin{split}
\mathcal{H}_{\text{fermion}}=&\mathcal{H}_{0}-\frac{1-g(r)\cos\Big(\hat{\alpha}(t)f(r)\Big)}{2r}\psi^{\dagger}_{f}(\vec{\sigma}\times\vec{\tau})\cdot\hat{r}\psi_{f}\\&+\frac{1}{2}\psi^{\dagger}_{f}\Big[\hat{\alpha}(t)f'(r)(\vec{\sigma}\cdot\hat{r})(\vec{\tau}\cdot\hat{r})+\frac{g(r)\sin\Big(\hat{\alpha}(t)f(r)\Big)}{r}\Big(\vec{\sigma}\cdot\vec{\tau}-(\vec{\sigma}\cdot\hat{r})(\vec{\tau}\cdot\hat{r})\Big)\Big] \psi_{f}\,,\label{eq:H4d}
\end{split}
\end{equation}
where we have used the two-component spinor notation for the Weyl fermions $\psi_{f}$. This Hamiltonian can be further simplified if we perform the following field redefinition for the fermions 
\begin{equation}
    \tilde{\psi}_{f}=U\psi_{f}\,,\quad \tilde{A}=A\,,\quad U=e^{-i\hat{\alpha}(t)f(r)\vec{T}\cdot\hat{r}}\,,\label{eq:globaltrans}
\end{equation}
which gives
\begin{equation}
\mathcal{H}_{\text{fermion}}=\tilde{\mathcal{H}}_{0}-\frac{1-g(r)}{2r}\tilde{\psi}^{\dagger}_{f}(\vec{\sigma}\times\vec{\tau})\cdot\hat{r}\tilde{\psi}_{f}+\frac{\partial_{t}\hat{\alpha}(t)}{2}f(r)\tilde{\psi}^{\dagger}_{f}\vec{\tau}\cdot\hat{r}\tilde{\psi}_{f}\,.\label{eq:tildeH4d}
\end{equation}
Before we proceed, we should note that the transformation in Equ.~(\ref{eq:globaltrans}) is not a gauge transformation, and it changes the asymptotic fields and boundary conditions. Thus, the descriptions of the scattering in these two frames will be different. We will see later that they are respectively the proposals in \cite{vanBeest:2023dbu} and \cite{Brennan:2023tae}.

\subsection{The Mode Decomposition and the s-wave Reduction}\label{sec:2dreduction}

Since both frames Equ.~(\ref{eq:H4d}) and Equ.~(\ref{eq:tildeH4d}) share the same conserved charges:
\begin{equation}
    \vec{J}=\vec{L}+\vec{S}+\vec{T}\,,
\end{equation}
their fermion mode decompositions are the same. Thus, we can start from one of them and then obtain the other by the transformation Equ.~(\ref{eq:globaltrans}).\footnote{Note that $[U,\vec{J}]=0$ as well.} For simplicity, we will start from the frame Equ.~(\ref{eq:tildeH4d}) which has a much simpler Hamiltonian to deal with. 

For convenience, let's define
\begin{equation}
    \vec{K}=\vec{S}+\vec{T}\,.
\end{equation}
With this definition, the Hamiltonian Equ.~(\ref{eq:tildeH4d}) can be written as
\begin{equation}
\begin{split}
    &i\vec{\sigma}\cdot\vec{\nabla}-\frac{1-g(r)}{2r}(\vec{\sigma}\times\vec{\tau})\cdot\hat{r}+\frac{\partial_{t}\hat{\alpha}(t)f(r)}{2}\tau^{r}\,,\\=&i\sigma^{r}(\partial_{r}-\frac{\vec{\sigma}\cdot\vec{L}}{r})-\frac{1-g(r)}{2ri}[\vec{K}^2,\sigma^{r}]+\frac{\partial_{t}\hat{\alpha}(t)f(r)}{2}\tau^{r}\,,\\=&i\sigma^{r}(\partial_{r}-\frac{(\vec{J}-\vec{T})^2-\vec{S}^2-\vec{L}^2}{r})-\frac{1-g(r)}{2ri}[\vec{K}^2,\sigma^{r}]+\frac{\partial_{t}\hat{\alpha}(t)f(r)}{2}\tau^{r}\,,\label{eq:Hdecom}
    \end{split}
\end{equation}
sandwiched between $\tilde{\psi}^{\dagger}_{f}$ and $\tilde{\psi}_{f}$. The maximal set of commutating operators is
\begin{equation}
(\vec{J}^2,\vec{T}^{2},\vec{S}^{2},\vec{K}^{2},K_{r},J_{z})\,,
\end{equation}
which fully determines the fermion wavefunction in the angular, spin and SU(2) gauge sectors. Since we already know that $\vec{S}^2=\vec{T}^2=\frac{3}{4}$ for the fermion, we decompose the 4d fermion $\tilde{\psi}_{f}(t,r,\Omega)$ into 2d modes as
\begin{equation}
    \tilde{\psi}_{f}(t,r,\Omega)=\sum_{j,m,k,k_r}\frac{1}{r}\tilde{\psi}_{f}^{jmkk_r}(t,r)\mathcal{Y}^{jm}_{kk_r}(\Omega)\,,
\end{equation}
where $m$ is the value of $J_{z}$ and $\mathcal{Y}^{jm}_{kk_r}$ is given by the Wigner D-function as
\begin{equation}
\mathcal{Y}^{jm}_{kk_r}=\sqrt{\frac{2j+1}{4\pi}}D^{j}_{mk_{r}}(\Omega)\ket{kk_{r}}\,.
\end{equation}
We want to focus on the s-wave, i.e. $j=m=0$, sector. We note that this sector is particularly simple as in the $j=0$ sector $m=0$ also projects $k_r$ to zero. This sector thus has the following orthonormal basis
\begin{equation}
    (k,k_r=s+q)=\{(0,0)\,,(1,0)\}\,.
\end{equation}
In this sector, the Hamiltonian Equ.~(\ref{eq:Hdecom}) simplifies to
\begin{equation}
\begin{split}
   i\sigma^{r}(\partial_{r}+\frac{\vec{K}^2}{r})-\frac{1-g(r)}{2ri}[\vec{K}^2,\sigma^{r}]+\frac{\partial_{t}\hat{\alpha}(t)f(r)}{2}\tau^{r}\,.\label{eq:Hdecom2}
    \end{split}
\end{equation}
Let's define the s-wave 2d fermion field as
\begin{equation}
    \tilde{\psi}_{f}(t,r)=\tilde{\psi}_{f}^{00kk_{r}}\ket{kk_{r}}\,.
\end{equation}
Hence, the reduced Hamiltonian becomes
\begin{equation}
    \tilde{\mathcal{H}}^{j=0}=\tilde{\psi}^{\dagger}_{f}(t,r)\Big[ i\sigma^{r}(\partial_{r}+\frac{\vec{K}^2-1}{r})-\frac{1-g(r)}{2ri}[\vec{K}^2,\sigma^{r}]+\frac{\partial_{t}\hat{\alpha}(t)f(r)}{2}\tau^{r}\Big]\tilde{\psi}_{f}(t,r)\,,
\end{equation}
and the fermion fields only contain the charged-spin locked modes with $s+q=0$. In this basis, we have
\begin{equation}
    \sigma^{r}=\begin{pmatrix}
0 & 1 \\
1 & 0
\end{pmatrix}\,,\quad \tau^{r}=\begin{pmatrix}
0 & -1 \\
-1 & 0
\end{pmatrix}\,.
\end{equation}
As a result, we have the Hamiltonian
\begin{equation}
    \tilde{\mathcal{H}}^{j=0}=\tilde{\psi}^{\dagger}_{f}(t,r)\begin{pmatrix}
0 & i\partial_{r}+i\frac{g(r)}{r}-\frac{\partial_{t}\hat{\alpha}(t)f(r)}{2} \\
i\partial_{r}-i\frac{g(r)}{r}-\frac{\partial_{t}\hat{\alpha}(t)f(r)}{2} & 0
\end{pmatrix}\tilde{\psi}_{f}(t,r)\,,
\end{equation}
Then we will always use the $\mathbf{\frac{1}{2}}_{\text{spin}}\otimes\mathbf{\frac{1}{2}}_{\text{charge}}$ basis
\begin{equation}
    \{\ket{\uparrow-},\ket{\downarrow+}\}\,,\label{eq:basis}
\end{equation}
for the fermion, and the spin and the charge are defined with respect to the radial direction.\footnote{This is the direction that the fermions parallelly or antiparallelly along and the preserved U(1) gauge symmetry corresponds to.} Thus, the Hamiltonian is better phrased as
\begin{equation}
    \tilde{\mathcal{H}}^{j=0}=\tilde{\psi}^{\dagger}_{f}(t,r)\begin{pmatrix}
i\partial_{r} -\frac{\partial_{t}\hat{\alpha}(t)f(r)}{2} &i\frac{g(r)}{r} \\
-i\frac{g(r)}{r} & -i\partial_{r} +\frac{\partial_{t}\hat{\alpha}(t)f(r)}{2}
\end{pmatrix}\tilde{\psi}_{f}(t,r)\,.\label{eq:Hd2dtilde}
\end{equation}
The corresponding reduced Hamiltonian in the untilded frame Equ.~(\ref{eq:H4d}) is
\begin{equation}
    \mathcal{H}^{j=0}=\psi^{\dagger}_{f}(t,r)\begin{pmatrix}
i\partial_{r} -\frac{\hat{\alpha}(t)f'(r)}{2} &i\frac{g(r)}{r}e^{-i\hat{\alpha}(t)f(r)} \\
-i\frac{g(r)}{r}e^{i\hat{\alpha}(t)f(r)}  & -i\partial_{r} -\frac{\hat{\alpha}(t)f'(r)}{2}
\end{pmatrix}\psi_{f}(t,r)\,.\label{eq:H2d}
\end{equation}

\section{The Fermion-Monopole Scattering in the Untilded Frame}\label{sec:untildedframe}
In this section, we study the fermion-monopole scattering in the untilded frame Equ.~(\ref{eq:H2d}). We first analyze the Heisenberg time evolution of the fermion field, introducing the fermion-rotor model \cite{Polchinski:1984uw}. Then we propose a dual description of the fermion-rotor system. This dual description is exactly solvable and the operator dictionary is highly non-local. We resolve the energy conservation puzzle by studying the spectral representation of the rotor vertex and show it has support at arbitrarily small energy. Finally, we carefully analyze the system Equ.~(\ref{eq:H2d}) and develop a direct method to compute multi-point anomalous correlators, without using  cluster decomposition principle. These results fill in gaps in the calculations in earlier papers \cite{Polchinski:1984uw,Loladze:2024ayk,Loladze:2025jsq}, and are consistent with a direct bosonized analysis in \cite{vanBeest:2023dbu}.

\subsection{What Happens to a Single Ingoing Fermion?}\label{sec:intoout}

To understand the fate of an ingoing fermion, or in other words what comes out if one sends in a single low energy fermion towards the monopole, let's analyze the Heisenberg time evolution of the fermion field $\psi_{f}(t,x)$.

The corresponding Lagrangian is
\begin{equation}
     \mathcal{L}^{j=0}=\psi^{\dagger}_{f}(t,r)\begin{pmatrix}
i\partial_{t}-i\partial_{r}+\frac{\hat{\alpha}(t)f'(r)}{2} &-i\frac{g(r)}{r}e^{-i\hat{\alpha}(t)f(r)} \\
i\frac{g(r)}{r}e^{i\hat{\alpha}(t)f(r)}  &i\partial_{t} +i\partial_{r} +\frac{\hat{\alpha}(t)f'(r)}{2}
\end{pmatrix}\psi_{f}(t,r)\,.\label{eq:L2d}
\end{equation}
Let's define
\begin{equation}
    \psi_{f}(t,r)=\begin{pmatrix} &\psi_{f,1}(t,r)\\&\psi_{f,2}(t,r)\label{eq:nicepara}
    \end{pmatrix}\,.
\end{equation}
Thus, the equation of motion from Equ.~(\ref{eq:L2d}), i.e. the operator's time Heisenberg evolution equation, is
\begin{equation}
\begin{split}
    (\partial_{t}-\partial_{r}-i\frac{\hat{\alpha}(t)f'(r)}{2})\psi_{f,1}(t,r)-\frac{g(r)}{r}e^{-i\hat{\alpha}(t)f(r)}\psi_{f,2}(t,r)&=0\,,\\(\partial_{t}+\partial_{r}-i\frac{\hat{\alpha}(t)f'(r)}{2})\psi_{f,2}(t,r)+\frac{g(r)}{r}e^{i\hat{\alpha}(t)f(r)}\psi_{f,1}(t,r)&=0\,.\label{eq:Heisenberg}
    \end{split}
\end{equation}
We are only interested in the modes with energy $E\ll m_{W}$. Let's analyze Equ.~(\ref{eq:Heisenberg}) in different regimes of $r>0$ and match the results:
\begin{itemize}
    \item \textbf{The near core regime $r\ll m_{W}^{-1}$}: 
    
    In this regime, we can ignore the time variation.    Thus, Equ.~(\ref{eq:Heisenberg}) implies
\begin{equation}
    \psi_{f}(t,r)\sim \tanh\frac{m_{W}r}{2}\begin{pmatrix}
        &e^{-i\frac{\hat{\alpha}(t)f(r)}{2}}\\&-e^{i\frac{\hat{\alpha}(t)f(r)}{2}}
    \end{pmatrix} C(t)\,.
\end{equation}

\item \textbf{The asymptotic regime $E^{-1}\ll r$}:
In this regime, we have
\begin{equation}
        f(r)\rightarrow1\,,\quad\,f'(r)\rightarrow0,\quad\text{and }g(r)\rightarrow0\,.
    \end{equation}
Thus, we have the free equation
\begin{equation}
\begin{split}
    (\partial_{t}-\partial_{r})\psi_{f,1}(t,r)&=0\,,\\(\partial_{t}+\partial_{r})\psi_{f,2}(t,r)&=0\,,\label{eq:Heisenbergr2}
    \end{split}
\end{equation}
which implies
\begin{equation}
    \psi_{f}(t,r)=\begin{pmatrix}
        F(t+r)\\G(t-r)
    \end{pmatrix}\,,
\end{equation}
where $F$ and $G$ are analytic functions in this regime. This is the Callan-Rubakov result that the incoming fermions can only be negative charged and the outgoing ones must be positive charged. 

    \item \textbf{The matching regime $m_{W}^{-1}\ll r\ll E^{-1}$}:

The results in the above two regimes should be matched in the intermediate regime $m_{W}^{-1}\ll r\ll E^{-1}$. In this regime, we have 
\begin{equation}
    F(t+r)\approx F(t)\,,\quad G(t-r)\approx G(t)\,,\quad f(r)=1\,,\quad\text{and }\tanh\frac{m_{W}r}{2}\rightarrow1\,.
\end{equation}
As a result, the matching gives
\begin{equation}
    F(t)=e^{-i\frac{\hat{\alpha}(t)}{2}}C(t)\,,\quad G(t)=-e^{i\frac{\hat{\alpha}(t)}{2}}C(t)\,.
\end{equation}
\end{itemize}
As a result, in the IR limit and away from the core, we have
\begin{equation}
    \psi_{f}(t,r)=\begin{pmatrix} &\psi_{f,1}(t+r)\\&-e^{i\hat{\alpha}(t-r)}\psi_{f,1}(t-r)  \end{pmatrix}\,.\label{eq:finalpsiuntilded}
\end{equation}
Thus, the ingoing negative charged fermion will be converted to an outgoing positive charged fermion with a rotor dressing.

\subsection{Reduction to Polchinski's Fermion-Rotor Model}\label{sec:Polchinski}

The result in the above subsection is in fact faithfully captured by a simpler model. This model is designed by Polchinski in \cite{Polchinski:1984uw} and recently analyzed in \cite{Loladze:2025jsq}. In this model, we drop the $g(r)$ terms in the Hamiltonian and unfold the $r$-coordinate from $r>0$ to $-\infty<x<\infty$. Thus, the incoming fermions live in $x<0$, the outgoing fermions live in $x>0$ and we have single component fermions $\psi_{f}(t,x)$. Meanwhile the function $f(r)$ should be extend as an odd function to reflect the charges of the fermions. 

The fermionic Lagrangian of this model is
\begin{equation}
    \mathcal{L}^{j=0}=\sum_{f=1}^{2N}\psi_{f}^{\dagger}(t,x)\Big(i\partial_{t}+i\partial_{x}+\frac{\hat{\alpha}(t)f'(x)}{2}\Big)\psi_{f}(t,x)\,.\label{eq:Polchinski}
\end{equation}
The equation of motion, i.e. the Heisenberg time evolution, is 
\begin{equation}
    \Big(i\partial_{t}+i\partial_{x}+\frac{\hat{\alpha}(t)f'(x)}{2}\Big)\psi_{f}(t,x)=0\,.
\end{equation}
In the IR limit, the function $f'(x)$ is zero outside the core regime. Thus, the solution of the above equation can be written as
\begin{equation}
    \psi_{f}(t,x)=e^{i\frac{\hat{\alpha}(t-x)}{2}f(x)}\psi^{s}_{f}(t-x)\,\label{eq:Polchinskiresult}
\end{equation}
where $\psi_{f}^{s}$ is a smooth function.

Therefore, using the fact that $f(x)$ goes to $1$ as $m_{W}x\rightarrow\infty$ and $-1$ as $m_{W}x\rightarrow-\infty$, we can see that Equ.~(\ref{eq:Polchinskiresult}) is consistent with Equ.~(\ref{eq:finalpsiuntilded}).\footnote{Note that the fermion's definition is ambiguous up to a constant phase. This explains the minus sign in the second component of Equ.~(\ref{eq:finalpsiuntilded}).} We will thus use this simpler model hereafter.

\subsection{A Dual Description}\label{sec:dual}

Here we provide a new dual description of the fermion-rotor system (\ref{eq:Polchinski}). First, we introduce an auxiliary decoupled scalar field $\Phi(t,x)$ on the half space $x>0$ with the Dirichlet boundary condition imposed at $x=0$. In Appendix.~\ref{sec:derivation}, We derive the following exact duality

\begingroup
\small
\begin{equation}
\begin{aligned}
\mathcal L_{\mathrm L}
 &= \int dx \sum_{f=1}^{2N} \psi_f^\dagger
    \bigl(i\partial_t + i\partial_x + q_f\,\alpha(t)\,f'(x)\bigr)\psi_f
    + \frac{I}{2}\dot\alpha^2 - H_{\mathrm{ct}}[\alpha] - \sum_{f=1}^{2N}\frac{q_f^2}{8\pi}
    \int_0^\infty dx\,\partial_\mu\Phi\,\partial^\mu\Phi\,, \\[6pt]
 &\quad\quad\quad\quad\quad\quad\quad\quad\quad\quad\quad\quad\quad\quad\quad\quad\quad\Big\Updownarrow\;\text{duality} \\[6pt]
\mathcal L_{\mathrm R}
 &= \int dx \sum_{f=1}^{2N} \Psi_f^\dagger
    \bigl(i\partial_t + i\partial_x\bigr)\Psi_f
    + \frac{I}{2}\dot\alpha^2 - \sum_{f=1}^{2N}\frac{q_f^2}{8\pi}
    \int_0^\infty dx\,\partial_\mu\phi\,\partial^\mu\phi
    + \eta(t)\bigl(\phi(t,0) - 2\alpha(t)\bigr)\,.
\end{aligned}
\label{eq:lagrangian-duality}
\end{equation}
\endgroup
We use $L$ and $R$ to denote the original formulation and the dual formulation, $\eta(t)$ is a Lagrange multiplier, the counter term is given in Equ.~(\ref{eq:momentumCT}) and we are in the IR limit where $f'(x)=2\delta(x)$. The operator dictionary is provided in Appendix.~\ref{sec:operatordic}.

This dual description is extremely simple as  it consists of two quadratic sectors. The first sector is $2N$ free fermions $\Psi_{f}(t,x)$ on the whole plane. The second sector is rotor together with one free boson $\phi(t,x)$ on the half plane $x\geq0$. The boundary value of boson $\phi$ is coupled to the rotor through Lagrange multiplier $\eta$. The dual description can be canonically quantized, and we will discuss its quantization and two implications below.

\subsubsection{Implication I: Conservation of Charge and Energy}\label{sec:conservation}
We first note that the single fermion scattering as described by Equ.~(\ref{eq:finalpsiuntilded}) and Equ.~(\ref{eq:Polchinskiresult}) obeys the electric charge conservation. Because as we can see from Equ.~(\ref{eq:basis}) and Equ.~(\ref{eq:L2d}), the ingoing fermion has charge $-\frac{e}{2}$ and the outgoing fermion has charge $\frac{e}{2}$ while the rotor vertex $e^{i\hat{\alpha}(t)}$ has charge $e$. Thus, for example in Equ.~(\ref{eq:Polchinskiresult}) we can consider the fermion field $\psi_{f}(t,x)$ just before ($x=0^{-}$) and just after the scattering ($x=0^{+}$). Let's define
\begin{equation}
    \psi_{f,\text{in}}(t)=\psi_{f}(t,0^{^{-}})\,,\quad\psi_{f,\text{out}}(t)=\psi_{f}(t,0^{^{+}})\,,\label{eq:inoutpsi}
\end{equation}
which implies that
\begin{equation}
    e^{-i\hat{\alpha}(t)}\psi_{f,\text{out}}(t)=\psi_{f,\text{in}}(t)\,.\label{eq:psibc}
\end{equation}
That is an incoming charge $-\frac{e}{2}$ fermion will be converted to an outgoing charge $\frac{e}{2}$ fermion, together with a rotor vertex whose charge is $-e$. This is nothing but the charge-depositing boundary condition that has been understood in the literature. However, the puzzle is that this charge-depositing effect means the rotor will be excited. The free rotor Hamiltonian is
\begin{equation}
    H_{\alpha,0}=\frac{p_{\alpha}^2}{2I}\,,
\end{equation}
which has a gap $\Delta_{0}=\frac{1}{2I}\sim e^2 m_{W}$ (remember that the free rotor is $2\pi$ periodic). Moreover, for the free rotor with the zero charge ground state $\ket{0}$ the state $V^{\dagger}(t)\ket{0}$ produced by the rotor vertex $V(t)=e^{i\alpha(t)}$ is an energy eigenstate with energy $\Delta_{0}$. Thus, it might be puzzling how a low energy fermion with energy $E\ll e^2 m_{W}$ could be converted to a fermion and a rotor vertex just after the scattering as described by Equ.~(\ref{eq:psibc}), or in other words how the energy could be conserved in this single fermion scattering process. The proposed answer for this puzzle in \cite{Loladze:2025jsq} is: \textit{ ... the resulting rotor must be highly off-shell and cannot last longer than a time scale $\sim I\ll1/E$.... any charge on the rotor quickly decays.}\footnote{See bottom of page 5 in the arxiv.org version of \cite{Loladze:2025jsq}. We thank John Terning for pointing out this problem to us.} Here, we will directly resolve this puzzle by showing that the Hamiltonian in the rotor sector is actually not gapped. This can be easily done using the dual description. We will canonically quantize the rotor and scalar boson sector, and see that the rotor vertex $e^{i\hat{\alpha}(t)}$ in fact has a continuous spectral representation.

For the sake of clarity to see how the discreteness of the rotor and scalar boson Hamiltonian emerges in the limit when they are decoupled, let's introduce a tunable parameter $0<g<2$ and consider the Lagrangian
\begin{equation}
    \mathcal{L}_{\alpha\phi}=-\int_0^\infty dx\,\frac{\gamma}{2}
\partial_\mu\phi(t,x)\partial^\mu\phi(t,x)+\eta(t)\bigl(\phi(t,0)-g\alpha(t)\bigr)+\frac{I}{2}\dot{\alpha}(t)^2\,,\quad \gamma\equiv\frac{\sum_{f=1}^{2N}q_{f}^2}{4\pi}\,.
\end{equation}
The Lagrange multiplier $\eta(t)$ imposes the constraint
\begin{equation}
    \phi(t,0)=g\alpha(t)\,.\label{eq:phibc}
\end{equation}
For the sake of convenience, let's take an IR cutoff $L$ such that the boson lives on $0<x<L$ and we impose the Neumann boundary condition
\begin{equation}
    \partial_{x}\phi(t,L)=0\,.\label{eq:IRbc}
\end{equation}
In addition to the boundary conditions Equ.~(\ref{eq:phibc}) and Equ.~(\ref{eq:IRbc}), we have the equations of motion
\begin{equation}
    (\partial_{t}^2-\partial_{x}^2)\phi(t,x)=0\,,\quad \gamma g\partial_{x}\phi(t,0)-I\ddot{\alpha}(t)=0\,.
\end{equation}
Let's do the modes decomposition of the scalar and the rotor
\begin{equation}
    \alpha(t)=\theta(t)+\frac{1}{g}\sum_{n}u_{n}(0)(\phi_{n}+\phi^{*}_{n})(t)\,,\quad \phi(t,x)=g\theta(t)+\sum_{n}u_{n}(x)(\phi_{n}+\phi_{n}^{*})(t)\,,\quad \phi_{n}(t)=\phi_{n}e^{-ik_{n}t}
\end{equation}
where we have a compact mode $\theta(t)=\theta(t)+2\pi$ and a tower of the IR regulated continuous modes, which are thus discrete, labeled by $n$. The above summation is only over $k_{n}>0$. We have thus the equations of motion and the boundary conditions
\begin{equation}
    (k_{n}^2+\partial_{x}^{2})u_{n}(x)=0\,,\quad \ddot{\theta}(t)=0\,,\quad\gamma g^{2}u'_{n}(0)=-Ik_{n}^{2}u_{n}(0)\,,\quad u'_{n}(L)=0\,.
\end{equation}
Thus, $\theta(t)$ is a compact (0+1)-d mode and we will have to deal with the canonical quantization of the other modes. The solution of the mode functions is
\begin{equation}
    u_{n}(x)=A_{n}\frac{Ik_{n}\sin k_{n}x-\gamma g^{2}\cos k_{n}x}{Ik_{n}}\,,
\end{equation}
and the quantization condition comes from
\begin{equation}
    \cos k_{n}L+\frac{\gamma g^{2}}{Ik_{n}}\sin k_{n}L=0\,.\label{eq:kquant}
\end{equation}
The coefficients $A_{n}$ are determined by the normalization condition (for $k_{n},k_{m}>0$)
\begin{equation}
    (u_{n}(x)e^{ik_{n}t},u_{m}(x)e^{ik_{m}t})_{\text{KG}}=\delta_{mn}\,,\quad (u_{n}(x)e^{ik_{n}t},\theta)_{\text{KG}}=0\,.
\end{equation}
The Klein-Gordon norm is
\begin{equation}
    (v,u)_{\text{KG}}=i\gamma\int_{0}^{L}dx \big(v^{*}\partial_{t}u-\partial_{t}v^{*}u\big)(t,x)+i\frac{I}{g^{2}}\big(v^{*}\partial_{t}u-\partial_{t}v^{*}u\big)(t,0)\,.
\end{equation}
Thus, it is easy to see that
\begin{equation}
    (u_{n}(x)e^{ik_{n}t},\theta(t))_{\text{KG}}=0\,,\quad  (u_{n}(x)e^{ik_{n}t},u_{m}(x)e^{ik_{m}t})_{\text{KG}}=\frac{\gamma |A_{n}|^2}{k_{n}}\Big[L(k_{n}^{2}+\frac{\gamma^2 g^{4}}{I^2})+\frac{\gamma g^{2}}{I}\Big]\delta_{mn}\,.
\end{equation}
Thus, we have for $k_{n}>0$
\begin{equation}
    A_{n}=\sqrt{\frac{k_{n}}{\gamma[L(k_{n}^{2}+\frac{\gamma^2 g^{4}}{I^2})+\frac{\gamma g^{2}}{I}\Big]}}\,.
\end{equation}
It is easy to work out the canonically quantized operators and they are
\begin{equation}
\begin{split}
    \hat{\phi}(t,x)&=g\hat{\theta}(t)+\sum_{n}u_{n}(x)(\hat{a}_{n}e^{ik_{n}t}+\hat{a}^{\dagger}_{n}e^{-ik_{n}t})\,,\\\hat{\alpha}(t)&=\hat{\theta}(t)+\frac{1}{g}\sum_{n}u_{n}(0)(\hat{a}_{n}e^{ik_{n}t}+\hat{a}^{\dagger}_{n}e^{-ik_{n}t})\,,\label{eq:quantcan}
    \end{split}
\end{equation}
where 
\begin{equation}
    [\hat{a}_{n},\hat{a}^{\dagger}_{m}]=\delta_{nm}\,,\quad[\hat{a}_{n},\hat{a}_{m}]=0\,,\label{eq:aa}
\end{equation}
and the sum is over $n$ for $k_{n}>0$. The canonically quantized Hamiltonian is
\begin{equation}
    \hat{H}_{\alpha\phi}=\frac{\hat{p}_{\theta}^2}{2I(1+I^{-1}\gamma g^{2}L)}+\sum_{n}k_{n}(\hat{a}^{\dagger}_{n}\hat{a}_{n}+\frac{1}{2})\,,\quad \hat{p}_{\theta}=(I+\gamma g^{2}L)\dot{\theta}(t)\,,\label{eq:canonicalHrotor}
\end{equation}
where we emphasize again that the sum is over $k_{n}>0$ and $\hat{p}_{\theta}$ is the canonical momentum of $\theta(t)$. Since $\theta(t)=\theta(t)+2\pi$, the first term in the Hamiltonian has a discrete spectrum and a gap.
\begin{equation}
    \Delta=\frac{1}{2I}\frac{1}{(1+I^{-1}\gamma g^{2}L)}\label{eq:gapnew}
\end{equation}

As a result, we can see that if we remove the IR regulator by setting the dimensionless constant $\frac{\gamma g^2L}{I}\rightarrow\infty$, the Hamiltonian in the rotor and scalar boson sector has a continuous spectrum. This is because the above gap $\Delta\ll\frac{1}{2I}$, the $k_{n}$ will be continuous, and the discrete sum $\sum_{n}$ and the canonical commutation relation Equ.~(\ref{eq:aa}) should be replaced by the continuous versions. While the discrete spectrum will be restored if we send the coupling $g$ to zero first.

Furthermore, the rotor vertex $V(t)=e^{i\hat{\alpha}(t)}$ has a spectral representation as follows
\begin{equation}
    C_{L}(t)\equiv\bra{\Omega}V^{\dagger}(t)V(0)\ket{\Omega}=e^{-\frac{it}{2I(1+I^{-1}\gamma g^2 L)}+\sum_{n}c_{n}^{2}(e^{-ik_{n}t}-1)}=\int_{0}^{\infty}dE\rho_{L}(E)e^{-iEt}\,,
\end{equation}
where
\begin{equation}
    \rho_{L}(E)=e^{-\sum_{n}c_{n}^{2}}\sum_{\{N_{n}\}}\Big[\prod_{n}\frac{(c_{n}^2)^{N_n}}{N_{n}!}\Big]\delta\big(E-\frac{1}{2I(1+I^{-1}\gamma g^2 L)}-\sum_{n}k_{n}N_{n}\big)\,,\quad c_{n}\equiv\frac{u_{n}(0)}{g}\,.
\end{equation}
Thus, in the continuous limit we have the quantization condition Equ.~(\ref{eq:kquant})
\begin{equation}
    k_{n}L+\arctan\frac{Ik_{n}}{\gamma g^{2}}=n\pi\,,\quad n=0,1,2,\cdots\,,
\end{equation}
implies
\begin{equation}
    \frac{dn}{dk}=\frac{1}{\pi}\Big[L+\frac{I^{-1}\gamma g^{2}}{k^2+(I^{-1}\gamma g^2)^2}\Big]\,.
\end{equation}
Thus, we have
\begin{equation}
    C(t)=\lim_{\frac{L\gamma g^2}{I}\rightarrow\infty}C_{L}(t)=e^{\int_{0}^{\infty}dk \frac{dn}{dk}c_{n}^2(e^{-ikt}-1)}=e^{\int_{0}^{\infty}d\omega\frac{1}{\pi\gamma g^2}\frac{1}{\omega\big(1+(\omega I\gamma^{-1}g^{2})^2\big)}(e^{-i\omega t}-1)}\,,
\end{equation}
and the low energy limit of $\rho(E)$
\begin{equation}
    \rho(E)\rightarrow\frac{e^{-\frac{\gamma_{E}}{\pi\gamma g^{2}}}}{\Gamma[\frac{1}{\pi\gamma g^{2}}](I^{-1}\gamma g^{2})^{\frac{1}{\pi\gamma g^{2}}}}E^{\frac{1}{\pi\gamma g^{2}}-1}\,,\quad \text{as } E\rightarrow0^{+}\,,
\end{equation}
where $\gamma_{E}$ is Euler's constant. Thus, the coupled rotor vertex has spectral weight at arbitrarily small positive energies. Low-energy scattering can therefore access the rotor vertex without paying a huge energy as the excitation gap of an isolated rotor.\footnote{See \cite{Hamada:2026jgj} for a similar conclusion using a different approach.}

\subsubsection{Implication II: Twisting Operators}
Having obtained the dual description and its exact quantization, we can compute various commutators between the operators in the origin description. 

An immediate commutator of importance is
\begin{equation}
    [\hat{\alpha}(t-x),\psi_{f}(t,y)]=[\hat{\alpha}(t-x),e^{A_{f,R}}\Psi_{f}(t,y)e^{-A_{f,R}^{\dagger}}]\,,
\end{equation}
where we have used the operator dictionary in Equ.~(\ref{eq:dictionary}). Let's first compute this commutator and then explain its physical importance. The operator $A_{f,R}$ is linear in the rotor and boson, so we just have to compute
\begin{equation}
    [\hat{\alpha}(t-x),(A_{f,R}-A^{\dagger}_{f,R})(t,y)]=-[\hat{\alpha}(t-x),\frac{i\pi q_{f}}{Q}\int_{-\infty}^{\infty}dz\text{sign}(z-y)B^{\text{can}}_{R}(t,z)]\,,
\end{equation}
where when we have a general parameter $g$ we have
\begin{equation}
    B_{R}^{\text{can}}(t,x)=\gamma \phi'(t,|x|)-\gamma\text{sign}(x)\dot{\phi}(t,|x|)+2\gamma g\alpha(t)\delta(x)\,.
\end{equation}
Thus, we can use the canonical quantized operators Equ.~(\ref{eq:quantcan}) and Equ.~(\ref{eq:aa}) to get
\begin{equation}
    [\hat{\alpha}(t-x),\psi_{f}(t,y)]=-\frac{q_{f}}{\gamma g}\Big[\Theta(y)\Theta(x-y)+\Theta(-y)\Theta(y-x)\Big]\big(1-e^{-\frac{\gamma g^{2}}{I}|x-y|}\big)\psi_{f}(t,y)\,.
\end{equation}
Let's then focus on the case on the right half plane $x,y>0$, i.e. the regime of outgoing fermions. Let's define
\begin{equation}
    \mathcal{T}(t,x)=e^{i\alpha(t-x)}\,,
\end{equation}
for which we have
\begin{equation}
    \mathcal{T}(t,x)\psi_{f}(t,y)=\begin{cases}\psi_{f}(t,y)\mathcal{T}(t,x)\,,\quad\text{for $y>x$}\,,\\e^{-i\frac{q_{f}}{\gamma g}(1-e^{-\frac{\gamma g^2}{I}(x-y)})}\psi_{f}(t,y)\mathcal{T}(t,x)\,,\quad\text{for $0<y<x$}\,.
    \end{cases}
\end{equation}
This result reproduces a corrected version of Equation (2.23) in \cite{Loladze:2025jsq},\footnote{We note that in fact our rotor $\alpha(t)$ is defined to be the opposite of their rotor. Therefore, the correct matching requires the twisted factor $e^{-\frac{4\pi i}{N}}$ in their (2.23) to be $e^{\frac{4\pi i}{N}}$. We believe this typo in \cite{Loladze:2025jsq} is propagated from their (2.20) where the correct last line should not have a minus sign.} if we take the IR limit $\gamma g^2\frac{x-y}{I}\gg1$ and consider all the fermions to be in the fundamental representation of the gauge $SU(2)$ and $g=2$. Therefore, according to \cite{Loladze:2025jsq}, the operator $\mathcal{T}(t,x)$ with $x>0$ is a twist operator with the branch cut from the monopole core to $x$. Moreover, from our result Equ.~(\ref{eq:Polchinskiresult}) the outgoing fermion has an additional phase $e^{i\alpha(t-x)}$ compared with the incoming fermion. Thus, the outgoing fermion is dressed by the twist operator $\mathcal{T}$ from the monopole core to the outgoing fermion. In other words, the outgoing fermion Hilbert space is twisted compared with the ingoing fermion Hilbert space. This is consistent with \cite{vanBeest:2023dbu,Loladze:2025jsq}.

\subsection{A Direct Computation of the Anomalous Correlators}\label{sec:correlators}
In this subsection, we focus on the original description of the fermion-rotor system. The non-anomalous fermion correlators can be computed using the generating function Equ.~(\ref{eq:Zpartial}). One can also consider the correlators of fermions and rotor vertices. This has been discussed in detail in \cite{Loladze:2025jsq}, with a few minor errors corrected by our Appendix. Thus, we won't discuss these correlators here. However, the anomaly induced fermion correlators were claimed to have a $0\times\infty$ ambiguity and were indirectly computed in \cite{Loladze:2025jsq} following \cite{Polchinski:1984uw} using the cluster decomposition of the non-anomalous correlators. In this section, we would like to directly compute these anomaly induced fermion correlators, i.e. anomalous correlators. In fact, the computation of these correlators conceptually follows a different approach than the non-anomalous one. This is because their origin is the fermion zero modes sourced by the rotor configurations that give a nonzero fermion number anomaly. 

Let's first explain the relationship between the correlators we are computing and the anomaly. We begin with an $SU(2)$ gauge theory coupled to $N_f=2N$ flavors of left-handed fermions in the fundamental representation in 4d. The total fermion-number current exhibits the chiral ABJ anomaly,
\begin{equation}
    \mathrm{d}\star j = \frac{N_f}{8\pi^2}\operatorname{tr}(F\wedge F),
    \qquad
    j^\mu=\sum_{f=1}^{N_f}\bar{\psi}_f\gamma^\mu\psi_f.
\end{equation}
Instantons are finite-action solutions of the Euclidean Yang–Mills equations. For an instanton of positive topological charge $Q$, the index theorem implies the existence of $QN_f$ fermion zero modes. These zero modes give rise to the fermion-number-violating correlators computed by ’t Hooft in \cite{tHooft:1976snw}
\begin{equation}
    \bigl\langle \underbrace{\psi\psi\cdots\psi}_{QN_f\ \mathrm{}}\bigr\rangle
    \propto e^{-S_{\mathrm{inst}}},
    \qquad
    S_{\mathrm{inst}}=\frac{8\pi^2Q}{g^2}.
\end{equation}
These are the anomalous correlators we are computing in this section. Such amplitudes are therefore exponentially suppressed at weak coupling for standard instanton configurations around the vacuum. However, in the monopole background, the anomaly equation reduces to
\begin{equation}
    (\partial_t+\partial_x)J(t,x)
    =\frac{N_f}{4\pi}\dot{\alpha}(t)f'(x),
\end{equation}
where $J(t,x)$ is the total chiral fermion-number density in the effective 2d theory. Physically, the anomaly is sourced by $F\wedge F\propto\vec{E}\cdot\vec{B}$: the monopole supplies a static magnetic field, while dynamics of the rotor generates an electric field proportional to $\dot{\alpha}(t)f'(x)$. Instanton topology in the microscopic theory is encoded by rotor winding in the effective description, suggesting that rotor winding configurations can likewise mediate fermion-number-violating correlators and encoding the same physics of ``Abelian Instanton" in \cite{Csaki:2024ajo}. We have shown in the previous section that the energy required to excite the charged rotor vertex can be arbitrarily small. Therefore, these processes may evade the exponential suppression associated with the ordinary Yang–Mills instantons. This expectation is consistent with Rubakov’s observation that the relevant action can be made arbitrarily small \cite{Rubakov:1982fp}.

We will compute the anomalous correlators in the Euclidean signature. We first compactify the Euclidean time $\tau=\tau+\beta$ and at the end take the $\beta\rightarrow\infty$ limit. The relationship between the Euclidean time and the Lorentzian time is
\begin{equation}
    t=-i\tau\,.
\end{equation}
The Euclidean action of the fermion-rotor system is
\begin{equation}
\begin{split}
    S_{E}=\int_{0}^{\beta} d\tau\Big[\int dx\,\sum_{f=1}^{2N}\psi_f^\dagger(\tau,x)
\bigl(\partial_\tau-i\partial_x
-q_f\alpha(\tau)f'(x)\bigr)\psi_f(\tau,x)
+\frac{I}{2}\dot\alpha(\tau)^2+H_{\mathrm{ct}}[\alpha]\Big]\,.
\end{split}
\end{equation}
The mode decomposition in the rotor sector with a winding number $w$ is
\begin{equation}
    \alpha(\tau)=\alpha_{0}+\frac{2\pi w\tau}{\beta}+\tilde{\alpha}(\tau)\,,\quad \tilde{\alpha}(\tau+\beta)=\tilde{\alpha}(\tau)\,,\quad\int_{0}^{\beta}d\tau\tilde{\alpha}(\tau)=0\,,\label{eq:rotordecomposition}
\end{equation}
where the zero mode $\alpha_{0}$ is the $2\pi$ periodic. The periodic part has the following mode expansion
\begin{equation}
    \tilde{\alpha}(\tau)=\sum_{n\neq 0}\alpha_{n}e^{\frac{2\pi in\tau }{\beta}}\,,\quad \alpha_{n}^{*}=\alpha_{-n}\,. \label{eq:rotordecomposition2}
\end{equation}

\subsubsection{The Necessity of the Winding Modes}
Before we study the effects of the winding configurations of the rotor  on the fermions, let's first explain why they must be included in the Euclidean path integral of the system. For this purpose, it is enough to look at one example computed by the Euclidean path integral and see the necessity of including the winding configurations to get the correct result. We consider the thermal partition function of the free rotor theory .
\begin{equation}
    Z_{\text{free rotor}}(\beta)=\Tr e^{-\beta H_{\alpha,0}}=\sum_{n=-\infty}^{\infty}e^{-\frac{n^2\beta}{2I}}\,,\label{eq:Zrotorfree}
\end{equation}
where we used the exact energy spectrum of a free rotor. The Euclidean path integral computes this quantity as
\begin{equation}
    Z_{\text{free rotor}}(\beta)=\int D\alpha e^{-\int_{0}^{\beta}d\tau\frac{I\dot{\alpha}(\tau)^2}{2}}\,.
\end{equation}
If we don't include the winding configurations, we will have
\begin{equation}
\begin{split}
    Z_{0}(\beta)&=\int Cd\alpha_{0} D\tilde{\alpha}e^{-\int_{0}^{\beta}d\tau\frac{\dot{\tilde{\alpha}}(\tau)^2I}{2}}=C\int d\alpha_{0}\prod_{n\neq0}d\alpha_{n}e^{-\sum_{n\neq0}\alpha_{n}\alpha_{-n}\frac{4\pi^2n^2I}{2\beta}}\,,\\&=2\pi\frac{1}{\sqrt{2}}\prod_{n>0}\frac{\beta}{4\pi In^2}=\sqrt{\frac{2\pi I}{\beta}}\,,
    \end{split}
\end{equation}
where we used $0<\alpha_{0}<2\pi$ and $C=\frac{1}{\sqrt{2}}$, which can be fixed by considering the free harmonic oscillator. However, the exact result Equ.~(\ref{eq:Zrotorfree}) can be written in a different form using the Poisson resummation formula as
\begin{equation}
Z_{\text{free rotor}}(\beta)=\sum_{k=-\infty}^{\infty}\sqrt{\frac{2\pi I}{\beta}}e^{-\frac{2\pi^2 Ik^2}{\beta}}\,.\label{eq:Zfreerotor2}
\end{equation}
Thus, to get a matched result, in the Euclidean path integral approach we have to sum over the winding sectors, i.e.
\begin{equation}
    Z_{\text{free rotor}}(\beta)=\sum_{w=-\infty}^{\infty}e^{-\frac{4\pi^2w^2I}{2\beta }}Z_{0}\,,
\end{equation}
which is now the same as Equ.~(\ref{eq:Zfreerotor2}) and thus Equ.~(\ref{eq:Zrotorfree}). As a result, in the Euclidean path integral computation of all the observables we must include the winding configurations of the rotor. For other physics application of using Poisson resummation formula to rewrite the sum of energy eigenstates into the sum of rotor winding modes, see \cite{Fan:2021ntg,Heidenreich:2020pkc}

\subsubsection{Fermion Zero Modes}
Having understood the necessity of the winding configurations of the rotor, let's study their effects on the fermion. The equation of motion of the fermion is
\begin{equation}
    \big(\partial_{\tau}-i\partial_{x}-q_{f}\alpha(\tau)f'(x)\big)\psi_{f}(\tau,x)=0\,.\label{eq:zeromodeeq}
\end{equation}
The solution has been understood in Sec.~\ref{sec:Polchinski}, and it is of the following form
\begin{equation}
    \psi_{f}^{0}(\tau,x)=e^{iq_{f}f(x)\alpha(\tau-ix)}u_{f}(\tau-ix)\,,
\end{equation}
where $u_{f}(\tau-ix)$ is smooth. Let's consider a winding configuration Equ.~(\ref{eq:rotordecomposition}) and Equ.~(\ref{eq:rotordecomposition2}),
which has a winding number $w$. The corresponding solution of Equ.~(\ref{eq:zeromodeeq}) is
\begin{equation}
    \psi_{f}^{0}(\tau,x)=e^{iq_{f}f(x)(\alpha_{0}+\sum_{n\neq0}\alpha_{n}e^{\frac{2\pi in(\tau-ix)}{\beta}})}e^{iq_{f}f(x)\frac{2\pi w(\tau-ix)}{\beta}}u_{f}(\tau-ix)\,,
\end{equation}
Let's give the fermion $\psi_{f}(\tau,x)$ the standard spin structure with anti-periodic boundary condition in $\tau$ and so we can expand $u_{f}(\tau-ix)$ as
\begin{equation}
    u_{f}(\tau-ix)=C(\tau-ix)\sum_{r}u_{r}e^{i\frac{2\pi (\tau-ix)}{\beta}r}\,,
\end{equation}
where we have $q_{f}=\frac{1}{2}$ so $r\in\mathbb{Z}\text{ }(\mathbb{Z}+\frac{1}{2})$ if $w$ is odd (even) and $C(\tau-ix)$ is a smooth function and is periodic in $\tau\rightarrow \tau+\beta$, which we will fix later. The fermion zero modes are normalizable solutions of Equ.~(\ref{eq:zeromodeeq}). Thus, for fermion zero modes, we must have
\begin{equation}
    \psi_{f}^{0}(\tau,x)\rightarrow0\,,\quad\text{as $x\rightarrow\pm\infty$}\,.
\end{equation}
This implies 
\begin{equation}
    C(\tau-ix)=e^{-iq_{f}(\sum_{n>0}\alpha_{n}e^{\frac{2\pi in(\tau-ix)}{\beta}}-\sum_{n<0}\alpha_{n}e^{\frac{2\pi in(\tau-ix)}{\beta}})}\,,
\end{equation}
and $u_{r}\neq0$ if $q_{f}w< r<-q_{f}w$. Thus, fermion zero modes exist iff $w<0$ and the number of fermion zero modes for the flavor $f$ obeying this condition is $-2q_{f}w$. In summary, for $-2q_{f}w\geq1$, we have the fermion zero modes
\begin{equation}
    \psi_{f,r}^{0}(\tau,x)=C(\tau-ix)e^{iq_{f}f(x)(\alpha_{0}+\sum_{n\neq0}\alpha_{n}e^{\frac{2\pi in(\tau-ix)}{\beta}})}e^{i\frac{2\pi(\tau-ix)}{\beta}r+iq_{f}f(x)\frac{2\pi w(\tau-ix)}{\beta}}\,.
\end{equation}
For convenience, let's define
\begin{equation}
    z\equiv e^{\frac{2\pi i(\tau-ix)}{\beta}}\,,
\end{equation}
and we have
\begin{equation}
    C(z)\equiv  e^{-iq_{f}(\sum_{n>0}\alpha_{n}z^{n}-\sum_{n<0}\alpha_{n}z^{n})}\,,
\end{equation}
Thus, the zero modes can be written as
\begin{equation}
    \psi_{f,r}^{0}(\tau,x)=C(z)e^{iq_{f}f(x)(\alpha_{0}+\sum_{n\neq0}\alpha_{n}z^{n})}z^{r}e^{iq_{f}f(x)\frac{2\pi w(\tau-ix)}{\beta}}\,.\label{eq:zeromodes}
\end{equation}
Before we proceed, we note that in the Euclidean signature the conjugate fermion $\psi_{f}^{\dagger}(\tau,x)$ obeys the equation of motion Equ.~(\ref{eq:zeromodeeq}) but with $\partial_{x}\rightarrow-\partial_{x}$. Thus, zero modes for $\psi_{f}^{\dagger}(\tau,x)$ requires $2q_{f}w\geq1$. For the sake of simplicity, we only focus on the cases that $-2q_{f}w\geq1$. Thus, there are no zero modes for the complex conjugate fermion $\psi_{f}^{\dagger}(\tau,x)$ as it requires $-2q_{f}w\leq-1$.

\subsubsection{Computing the Anomalous Correlators}\label{sec:anomalouscorrelators}
Now we are ready to compute the anomalous correlators. 

As an explicit example, let's first consider $N_{f}=2N=2$ and $q_{1}=q_{2}=\frac{1}{2}$. In this case, the lowest anomalous correlator is the time-ordered two-point function
\begin{equation}
    \langle T\psi_{1}(\tau_{1},x_{1})\psi_{2}(\tau_{2},x_{2})\rangle=\frac{1}{Z}\int D\alpha\prod_{f=1}^{2}D\psi^{\dagger}_{f}D\psi_{f}\psi_{1}(\tau_{1},x_{1})\psi_{2}(\tau_{2},x_{2})e^{-S_{E}[\psi^{\dagger},\psi,\alpha]}\,,
\end{equation}
for which the only contribution is from the winding configuration with $w=-1$. The path integral can be computed mode by mode. For the winding configuration with $w=-1$, there is one zero mode for each flavor. Thus, we have the fermion mode decomposition
\begin{equation}
    \psi_{f}(\tau,x)=\eta_{f}^{0}\psi_{f,0}^{0}(\tau,x)+\underbrace{\cdots\cdots}_{\text{nonzero modes}}\,,
\end{equation}
where all the mode coefficients , e.g. $\eta_{f}^{0}$, are Grassmann numbers. The mode decomposition of the rotor is Equ.~(\ref{eq:rotordecomposition}) with $w=-1$. The fermion path integral measure $D\psi_{f}$ contains the zero mode piece $d\eta_{f}^{0}$. Thus, we have
\begin{equation}
\begin{split}
    &\langle T\psi_{1}(\tau_{1},x_{1})\psi_{2}(\tau_{2},x_{2})\rangle\\&=\frac{\det^{2}(i\partial_{\tau}+\partial_{x})}{Z}\int Cd\alpha_{0}\prod_{n\neq0}d\alpha_{n}d\eta_{1}^{0}d\eta_{2}^{(0)} \mu\mathcal{N}_{2}(\beta)\psi_{1}^{0}(\tau_{1},x_{1})\psi_{2}^{0}(\tau_{2},x_{2})e^{-\sum_{n=1}^{\infty}(\frac{4\pi^2 I}{\beta}n^2+n)\alpha_{n}\alpha_{-n}-S_{w=-1}}\,,\label{eq:2pt}
    \end{split}
\end{equation}
where we have integrated out the nonzero modes of the fermion following the method in Appendix.~\ref{sec:derivation}, $\mu$ is a scale that we have to introduce to get the dimension of the fermions correct, $\mathcal{N}_{2}(\beta)=-\frac{e^{i\varphi_{2}}}{\beta\mu}$ is the normalization factor which we explain in Appendix.~\ref{sec:normalization}, and $\psi_{f}^{0}(\tau_{i},x_{i})$ denotes the zero mode part of the mode expansion of $\psi_{f}(\tau_{i},x_{i})$. The resulting integral in Equ.~(\ref{eq:2pt}) can be done piece by piece. First, the rotor zero mode part is
\begin{equation}
    \frac{C}{2\pi C}\int_{0}^{2\pi}d\alpha_{0}e^{i\alpha_{0}\frac{1}{2}\big(f(x_{1})+f(x_{2})\big)}=\delta_{\big(f(x_{1})+f(x_{2})\big)}\equiv\delta_{\sum_{i}\sigma_{i},0}\,.
\end{equation}
Second, the fermion zero mode part gives
\begin{equation}
    -\psi_{1,0}^{0}(\vec{x}_{1})\psi_{2,0}^{0}(\vec{x}_{2})\,,
\end{equation}
where, for the clarity of notation, we introduced $\vec{x}_{i}=(\tau_{i},x_{i})$. Third, we have the integrals of other rotor modes of the following general form
\begin{equation}
\begin{split}
    &\big(\prod_{n>0}\frac{\pi\beta}{4\pi^2 n^2 I+n\beta}\big)^{-1}\int \prod_{n>0}d\alpha_{n}d\alpha_{-n}z_{1}^{-\frac{f(x_{1})}{2}}z_{2}^{-\frac{f(x_{2})}{2}}\\&\quad\quad\quad\quad\quad\quad\quad\quad\quad\quad\times C(z_{1})C(z_{2}) e^{\frac{i}{2}\sum_{n\neq0}\alpha_{n}\big(f(x_{1})z_{1}^{n}+f(x_{2})z_{2}^{n}\big)-\sum_{n=1}^{\infty}(\frac{4\pi^2 I}{\beta}n^2+n)\alpha_{n}\alpha_{-n}}\,,\\&=z_{1}^{-\frac{f(x_{1})}{2}}z_{2}^{-\frac{f(x_{2})}{2}}e^{-\frac{1}{4}\sum_{n=1}^{\infty}\frac{\sum_{i=1}^{2}\big(f(x_{i})-1\big)z_{i}^{n}\sum_{j=1}^{2}\big(f(x_{j})+1\big)z_{j}^{-n}}{\frac{4\pi^2 In^2}{\beta}+n}}\,.\label{eq:generalform2pt}
    \end{split}
\end{equation}
This result can be simplified by defining
\begin{equation}
    L_{n}\equiv\sum_{x_i<0}z_{i}^{n}\,,\quad R_{n}\equiv\sum_{x_i>0}z_{i}^{-n}\,,
\end{equation}
which implies Equ.~(\ref{eq:generalform2pt}) equals
\begin{equation}
    \prod_{i=1}^{2}z_{i}^{-\frac{f(x_{i})}{2}}e^{\sum_{n=1}^{\infty}\frac{L_{n}R_{n}}{\frac{4\pi^2 In^2}{\beta}+n}}\,.
\end{equation}
Fourth, we have the weight from the winding mode action
\begin{equation}
    e^{-S_{w=-1}=}e^{-\frac{2\pi^2 I}{\beta}}\,.
\end{equation}
Now putting everything together and remember that $Z$ also has a factor $\det^{2}(i\partial_{\tau}+\partial_{x})$, we have
\begin{equation}
\langle T\psi_{1}(\tau_{1},x_{1})\psi_{2}(\tau_{2},x_{2})\rangle=-\mu\mathcal{N}_{2}(\beta)\delta_{\sum_{i}\sigma_{i},0}
z_{1}^{-\frac{f(x_{1})}{2}}z_{2}^{-\frac{f(x_{2})}{2}}e^{-\frac{2\pi^2 I}{\beta}+\sum_{n=1}^{\infty}\frac{L_{n}R_{n}}{\frac{4\pi^2 In^2}{\beta}+n}}\,.
\end{equation}
In the IR limit, we have $\frac{I}{\beta}\ll1$, $\frac{\abs{\vec{x}_{i}-\vec{x}_{j}}}{I}\gg1$ and $f(x)=\text{sign}(x)$. Therefore, the two-point function becomes
\begin{equation}
    \begin{split}
&\langle T\psi_{1}(\tau_{1},x_{1})\psi_{2}(\tau_{2},x_{2})\rangle\,,\\&=-\mu\mathcal{N}_{2}(\beta)\delta_{\sum_{i}\sigma_{i},0}z_{1}^{-\frac{f(x_{1})}{2}}z_{2}^{-\frac{f(x_{2})}{2}}e^{\sum_{n=1}^{\infty}\frac{L_{n}R_{n}}{n}}\,,\\&=-\mu\mathcal{N}_{2}(\beta)\delta_{\sum_{i}\sigma_{i},0}z_{1}^{-\frac{f(x_{1})}{2}}z_{2}^{-\frac{f(x_{2})}{2}}e^{-\sum_{x_l<0,x_s>0}\log(1-z_{l}/z_{s})}\,,\\&=-\mu\mathcal{N}_{2}(\beta)\delta_{\sum_{i}\sigma_{i},0}z_{1}^{-\frac{f(x_{1})}{2}}z_{2}^{-\frac{f(x_{2})}{2}}\prod_{x_l<0,x_s>0}\frac{z^{s}}{z_{s}-z_{l}}\,.
    \end{split}
\end{equation}
Furthermore, the IR limit also imposes $\frac{|\vec{x}_{i}-\vec{x}_{j}|}{\beta}\ll1$. Thus, we finally have
\begin{equation}
\langle T\psi_{1}(\tau_{1},x_{1})\psi_{2}(\tau_{2},x_{2})\rangle=e^{i\varphi_{2}}\frac{\delta_{\sum_{i}\sigma_{i},0}}{2\pi}\frac{1}{s_{ls}}\,,
\end{equation}
where $s_{ij}\equiv(i\tau_{i}-i\tau_{j}+x_{i}-x_{j})$, $x_{s}<0$ and $x_{l}>0$.

As a second example in the case with $N_{f}=2N=2$ and $q_{1}=q_{2}=\frac{1}{2}$, we can also compute the four-point function
\begin{equation}
    \begin{split}
        &\langle T\psi_{1}(\tau_{1},x_{1})\psi_{1}(\tau_{2},x_{2})\psi_{2}(\tau_{3},x_{3})\psi_{2}(\tau_{4},x_{4})\rangle\\&=\frac{1}{Z}\int D\alpha\prod_{f=1}^{2}D\psi^{\dagger}_{f}D\psi_{f}\psi_{1}(\tau_{1},x_{1})\psi_{1}(\tau_{2},x_{2})\psi_{2}(\tau_{3},x_{3})\psi_{2}(\tau_{4},x_{4})e^{-S_{E}[\psi^{\dagger},\psi,\alpha]}\,,
    \end{split}
\end{equation}
for which the only contribution is from the winding configuration with $w=-2$. Again, the path integral can be computed mode by mode. The fermion mode decomposition is
\begin{equation}
    \psi_{f}(\tau,x)=\eta^{(-1)}_{f}\psi_{f,-\frac{1}{2}}^{0}(\tau,x)+\eta^{(0)}_{f}\psi_{f,\frac{1}{2}}^{0}(\tau,x)+\underbrace{\cdots\cdots}_{\text{nonzero modes}}\,,
\end{equation}
The mode decomposition of the rotor is Equ.~(\ref{eq:rotordecomposition}) with $w=-2$. Thus, we have
\begin{equation}
\begin{split}
    \langle T\psi_{1}(\tau_{1},x_{1})&\psi_{1}(\tau_{2},x_{2})\psi_{2}(\tau_{3},x_{3})\psi_{2}(\tau_{4},x_{4})\rangle=\frac{\det^{2}(i\partial_{\tau}+\partial_{x})}{Z}\int Cd\alpha_{0}\prod_{n\neq0}d\alpha_{n}d\eta_{1}^{(-1)}d\eta_{1}^{(0)}d\eta_{2}^{(-1)}d\eta_{2}^{(0)}\\&\times \mu^{2}\mathcal{N}_{4}(\beta)\psi_{1}^{0}(\tau_{1},x_{1})\psi_{1}^{0}(\tau_{2},x_{2})\psi_{2}^{0}(\tau_{3},x_{3})\psi_{2}^{0}(\tau_{4},x_{4})e^{-\sum_{n=1}^{\infty}(\frac{4\pi^2 I}{\beta}n^2+n)\alpha_{n}\alpha_{-n}-S_{w=-2}}\,,\label{eq:4pt}
    \end{split}
\end{equation}
where, as in the 2pt calculation, we have integrated out the nonzero modes of the fermion following the method in Appendix.~\ref{sec:derivation}, $\mu$ is a scale that we have to introduce to get the dimension of the fermions correct, $\mathcal{N}_{4}(\beta)=\frac{e^{i\varphi_{4}}}{\beta^2\mu^2}$ is the normalization factor which we explain in Appendix.~\ref{sec:normalization}, and $\psi_{f}^{0}(\tau_{i},x_{i})$ denotes the zero mode part of the mode expansion of $\psi_{f}(\tau_{i},x_{i})$. The resulting integral in Equ.~(\ref{eq:4pt}) can be done piece by piece as in two-point function. First, the rotor zero mode part is
\begin{equation}
    \frac{C}{2\pi C}\int_{0}^{2\pi}d\alpha_{0}e^{i\alpha_{0}\frac{1}{2}\big(f(x_{1})+f(x_{2})+f(x_{3})+f(x_{4})\big)}=\delta_{\big(f(x_{1})+f(x_{2})+f(x_{3})+f(x_{4})\big)}\equiv\delta_{\sum_{i}\sigma_{i},0}\,.
\end{equation}
Second, the fermion zero mode part gives
\begin{equation}
    \big(\psi_{1,-\frac{1}{2}}^{0}(\vec{x}_{1})\psi_{1,\frac{1}{2}}^{0}(\vec{x}_{2})-\psi_{1,\frac{1}{2}}^{0}(\vec{x}_{1})\psi_{1,-\frac{1}{2}}^{0}(\vec{x}_{2})\big)\big(\psi_{2,-\frac{1}{2}}^{0}(\vec{x}_{3})\psi_{2,\frac{1}{2}}^{0}(\vec{x}_{4})-\psi_{2,\frac{1}{2}}^{0}(\vec{x}_{3})\psi_{2,-\frac{1}{2}}^{0}(\vec{x}_{4})\big)\,,
\end{equation}
where, for the clarity of notation, we introduced $\vec{x}_{i}=(\tau_{i},x_{i})$. Third, we have the integrals of other rotor modes of the following general form
\begin{equation}
\begin{split}
    &\big(\prod_{n>0}\frac{\pi\beta}{4\pi^2 n^2 I+n\beta}\big)^{-1}\int \prod_{n>0}d\alpha_{n}d\alpha_{-n}z_{1}^{r_{1}-f(x_{1})}z_{2}^{r_{2}-f(x_{2})}z_{3}^{r_{3}-f(x_{3})}z_{4}^{r_{4}-f(x_{4})}\\&\times C(z_{1})C(z_{2})C(z_{3})C(z_{4}) e^{\frac{i}{2}\sum_{n\neq0}\alpha_{n}\big(f(x_{1})z_{1}^{n}+f(x_{2})z_{2}^{n}+f(x_{3})z_{3}^{n}+f(x_{4})z_{4}^{n}\big)-\sum_{n=1}^{\infty}(\frac{4\pi^2 I}{\beta}n^2+n)\alpha_{n}\alpha_{-n}}\,,\\&=z_{1}^{r_{1}-f(x_{1})}z_{2}^{r_{2}-f(x_{2})}z_{3}^{r_{3}-f(x_{3})}z_{4}^{r_{4}-f(x_{4})}e^{-\frac{1}{4}\sum_{n=1}^{\infty}\frac{\sum_{i=1}^{4}\big(f(x_{i})-1\big)z_{i}^{n}\sum_{j=1}^{4}\big(f(x_{j})+1\big)z_{j}^{-n}}{\frac{4\pi^2 In^2}{\beta}+n}}\,.\label{eq:generalform}
    \end{split}
\end{equation}
This result can be simplified by defining
\begin{equation}
    L_{n}\equiv\sum_{x_i<0}z_{i}^{n}\,,\quad R_{n}\equiv\sum_{x_i>0}z_{i}^{-n}\,,
\end{equation}
which implies Equ.~(\ref{eq:generalform}) equals
\begin{equation}
    \prod_{i=1}^{4}z_{i}^{r_{i}-f(x_{i})}e^{\sum_{n=1}^{\infty}\frac{L_{n}R_{n}}{\frac{4\pi^2 In^2}{\beta}+n}}
\end{equation}
Fourth, we have the weight from the winding mode action
\begin{equation}
    e^{-S_{w=-2}=}e^{-\frac{8\pi^2 I}{\beta}}\,.
\end{equation}
Now putting everything together and remember that $Z$ also has a factor $\det^{2}(i\partial_{\tau}+\partial_{x})$, we have
\begin{equation}
    \begin{split}
&\langle T\psi_{1}(\tau_{1},x_{1})\psi_{1}(\tau_{2},x_{2})\psi_{2}(\tau_{3},x_{3})\psi_{2}(\tau_{4},x_{4})\rangle\,,\\&=\mu^{2}\mathcal{N}_{4}(\beta)\delta_{\sum_{i}\sigma_{i},0}(z_{1}-z_{2})(z_{3}-z_{4})\prod_{i=1}^{4}z_{i}^{-\frac{1}{2}-f(x_{i})}e^{-\frac{8\pi^2 I}{\beta}+\sum_{n=1}^{\infty}\frac{L_{n}R_{n}}{\frac{4\pi^2 In^2}{\beta}+n}}\,,
    \end{split}
\end{equation}
Therefore, in the IR limit, the four-point function becomes
\begin{equation}
    \begin{split}
&\langle T\psi_{1}(\tau_{1},x_{1})\psi_{1}(\tau_{2},x_{2})\psi_{2}(\tau_{3},x_{3})\psi_{2}(\tau_{4},x_{4})\rangle\,,\\&=\mu^{2}\mathcal{N}_{4}(\beta)\delta_{\sum_{i}\sigma_{i},0}(z_{1}-z_{2})(z_{3}-z_{4})\prod_{i=1}^{4}z_{i}^{-\frac{1}{2}-f(x_{i})}e^{\sum_{n=1}^{\infty}\frac{L_{n}R_{n}}{n}}\,,\\&=\mu^{2}\mathcal{N}_{4}(\beta)\delta_{\sum_{i}\sigma_{i},0}(z_{1}-z_{2})(z_{3}-z_{4})\prod_{i=1}^{4}z_{i}^{-\frac{1}{2}-f(x_{i})}e^{-\sum_{x_l<0,x_s>0}\log(1-z_{l}/z_{s})}\,,\\&=\mu^{2}\mathcal{N}_{4}(\beta)\delta_{\sum_{i}\sigma_{i},0}(z_{1}-z_{2})(z_{3}-z_{4})\prod_{i=1}^{4}z_{i}^{\frac{1}{2}}\prod_{x_l<0,x_s>0}\frac{1}{z_{s}-z_{l}}\,.
    \end{split}
\end{equation}
Furthermore, the IR limit also imposes $\frac{|\vec{x}_{i}-\vec{x}_{j}|}{\beta}\ll1$. Thus, we finally have
\begin{equation}
    \begin{split}
&\langle T\psi_{1}(\tau_{1},x_{1})\psi_{1}(\tau_{2},x_{2})\psi_{2}(\tau_{3},x_{3})\psi_{2}(\tau_{4},x_{4})\rangle\,,\\&=e^{i\varphi_{4}}\frac{\delta_{\sum_{i}\sigma_{i},0}}{(2\pi)^2}\frac{s_{12}s_{34}}{\prod_{x_{l}<0,x_{s}>0}s_{sl}}\,,
    \end{split}
\end{equation}
where $s_{ij}\equiv(i\tau_{i}-i\tau_{j}+x_{i}-x_{j})$.

The above method and results are easily generalized to higher-point correlators and also to cases of more flavors. For example, for the eight-point function in $q_{f}=\frac{1}{2}$ and $N_{f}=2N=2$ case, we have
\begin{equation}
     \begin{split}
&\langle T\psi_{1}(\tau_{1},x_{1})\psi_{1}(\tau_{2},x_{2})\psi_{1}(\tau_{3},x_{3})\psi_{1}(\tau_{4},x_{4})\psi_{2}(\tau_{5},x_{5})\psi_{2}(\tau_{6},x_{6})\psi_{2}(\tau_{7},x_{7})\psi_{2}(\tau_{8},x_{8})\rangle\,,\\&=e^{i\varphi_{8}}\frac{\delta_{\sum_{i}\sigma_{i},0}}{(2\pi)^4}\frac{\prod_{1\leq a<b\leq4}s_{ab}\prod_{5\leq c<d\leq8}s_{cd}}{\prod_{x_{l}<0,x_{s}>0}s_{sl}}\,,
    \end{split}
\end{equation}
where $\phi_{8}$ is an undetermined real number.

In summary, we have developed the general method to compute anomalous fermion correlators in the fermion-rotor system. 

\section{The Fermion-Monopole Scattering in the Tilded Frame}\label{sec:thetachanging}
Having understood the physics in the untilded frame, let's try to understand the tilded frame in Equ.~(\ref{eq:Hd2dtilde}). Since these two frames are related by a simple transform in Equ.~(\ref{eq:globaltrans}), all the physics in the untilded frame can be straightforwardly mapped to the tilded frame. Therefore, we will be brief and only focus on one problem that the untilded frame makes the answer easier to see. This is the energy conservation puzzle that we discussed in Sec.~\ref{sec:conservation}.

Instead of directly discussing Equ.~(\ref{eq:Hd2dtilde}), we will start with Equ.~(\ref{eq:Polchinski}) and transform it to the untilded frame using Equ.~(\ref{eq:globaltrans}). Since, with the counter term Equ.~(\ref{eq:momentumCT}), this is not an anomalous transformation, there wouldn't be any additional terms in the Lagrangian generated quantum mechanically. Thus, the Lagrangian is simply the classical transformation of Equ.~(\ref{eq:Polchinski}) under Equ.~(\ref{eq:globaltrans}) together with the rotor piece
\begin{equation}
\mathcal{L}^{j=0}=\tilde{\mathcal{L}}^{j=0}=\sum_{f=1}^{2N}\tilde{\psi}_{f}^{\dagger}(t,x)\Big(i\partial_{t}+i\partial_{x}-\frac{\partial_{t}\hat{\alpha}(t)f(x)}{2}\Big)\tilde{\psi}_{f}(t,x)+\frac{I}{2}(\partial_{t}\hat{\alpha}(t))^2\,.\label{eq:Polchinskitilde}
\end{equation}
Let's work out the canonical Hamiltonian associated with Equ.~(\ref{eq:Polchinskitilde}) and see that it has no energy puzzle. First, we have the canonical momenta
\begin{equation}
    \pi_{\tilde{\psi}_{f}}(t,x)=i\tilde{\psi}^{\dagger}_{f}(t,x)\,,\quad \Pi_{\alpha}(t)=I\partial_{t}\hat{\alpha}(t)-\frac{1}{2}\int dxf(x)\sum_{f=1}^{2N}\tilde{\psi}^{\dagger}_{f}\tilde{\psi}_{f}(t,x)=-i\frac{\partial}{\partial\alpha}\Big|_{\tilde{\psi}_{f}}\,.
\end{equation}
In terms of the canonical momenta, we have the canonical Hamiltonian
\begin{equation}
   H=\tilde{H}=-\int dx\sum_{f=1}^{2N}\tilde{\psi}^{\dagger}_{f}i\partial_{x}\tilde{\psi}_{f}(t,x)+\frac{\Big(\Pi_{\alpha}(t)+\frac{1}{2}\int dxf(x)\sum_{f=1}^{2N}\tilde{\psi}^{\dagger}_{f}\tilde{\psi}_{f}(t,x)\Big)^2}{2I}\,.\label{eq:Htheta}
\end{equation}
Now we can translate Equ.~(\ref{eq:inoutpsi}) into $\tilde{\psi}_{f}(t,x)$
\begin{equation}
    \psi_{f,\text{in}}(t)=e^{-i\frac{\hat{\alpha}(t)}{2}}\tilde{\psi}_{f}(t,0^{-})\,,\quad \psi_{f,\text{out}}(t)=e^{i\frac{\hat{\alpha}(t)}{2}}\tilde{\psi}_{f}(t,0^{+})\,.
\end{equation}
Therefore, we can see that the change in $
\Pi_{\alpha}$ is canceled by the change in $\frac{1}{2}\int dxf(x)\sum_{f=1}^{2N}\tilde{\psi}^{\dagger}_{f}\tilde{\psi}_{f}(t,x)$. Hence, there wouldn't be any $\frac{1}{2I}$ energy shift before and after the scattering. As a result, the energy puzzle is gone.

\section{Comments on Various Existing Proposals}\label{sec:comments}

There are various existing proposals for the massless fermion-monopole scattering in the recent literature \cite{Hamada:2022eiv,vanBeest:2023dbu,vanBeest:2023mbs,Brennan:2023tae,Csaki:2024ajo,Loladze:2024ayk,Loladze:2025jsq} regarding both how the outgoing fermions look and why the energy is conserved.\footnote{See also \cite{Bogojevic:2024xtx,Hamada:2026jgj}.} These papers made different proposals and used different approaches, and these proposals are not manifestly consistent with each other.  In this section, we will discuss these proposals and try to provide a unified picture based on the results of our analysis. We argue they are mutually consistent with each other and there is no intrinsic conflict between them. 

In \cite{Hamada:2022eiv,vanBeest:2023dbu,vanBeest:2023mbs,Loladze:2024ayk}, the author proposed that the Hilbert space of the ingoing fermions is different from the Hilbert space of the outgoing fermions. More precisely, in the IR limit, they are relatively twisted. In \cite{vanBeest:2023dbu,vanBeest:2023mbs}, this is found by first bosonizing the 2d folded fermions and then imposing the boundary condition for the bosons that preserves the relevant symmetries in the original fermion-monopole scattering. This approach is agnostic to the rotor. In fermion-rotor language,  \cite{Loladze:2024ayk,Loladze:2025jsq} noticed that a non-vanishing two-point correlator between an incoming fermion and an outgoing fermion with the same flavor requires an additional insertion of the rotor vertex $e^{i\alpha(t-x)}$, and arguing that this vertex is actually a twist operator for fermions. However, a puzzle exists regarding this rotor vertex. This is the energy conservation puzzle we discussed in Sec.~\ref{sec:conservation}. In our Sec.~\ref{sec:intoout}, we uncovered the above relative twisting between the ingoing and outgoing fermions, i.e. dressing by the rotor vertex $e^{i\alpha(t-x)}$, by directly solving the Heisenberg evolution of the fermion field operator in the fermion-rotor system we obtained from a careful $s$-wave reduction of the original 4d fermion-monopole system. We confirm that this result persists in Polchinksi's fermion-rotor model in Sec.~\ref{sec:Polchinski}. We also derived an exact dual description of Polchinski's fermion-rotor system (together with a decoupled free scalar sector). This dual description is very simple as it is a free system. It provides a direct derivation that the rotor vertex is a twist operator for fermions and an explicit resolution of the energy conservation puzzle. In \cite{Hamada:2022eiv}, the twist is obtained by treating the rotor classically, the Hilbert space of fermion will depend on the value of $\alpha$. Scattering process shifts $\alpha$ and thus twists the Hilbert space of the fermion. We generalized the result of \cite{Hamada:2022eiv} by providing a fully quantum mechanical treatment of the rotor in the dual description.

In \cite{Csaki:2024ajo}, the authors computed the anomalous fermion correlators by considering 4d fermions coupled with dynamical photons in the monopole background, and argued that in the 2d reduction of this system certain configurations of the 2d photon field gives a nonzero 4d instanton number.\footnote{It is not so clear if these configurations are on-shell.} Thus, there will be fermion zero modes sourced by such configurations and thus they will contribute to the path integral computation of the anomalous fermion correlators. The authors in \cite{Csaki:2024ajo} acknowledged that such correlators are captured by the bosonized descriptions in \cite{vanBeest:2023dbu,vanBeest:2023mbs} but claims that they did't have an instanton origin. In fermion-rotor model, anomalous fermion correlators are computed by cluster decomposition of the non-anomalous fermion correlators \cite{Polchinski:1984uw,Loladze:2025jsq}. In Sec.~\ref{sec:correlators}, we filled this gap by providing a direct computation of these anomalous fermion correlators and found that its origin is indeed instantonic. More specifically, we found that the winding configuration of the rotor sources fermion zero modes which gives non-zero anomalous fermion correlators. We also showed that such winding configurations of the rotor must be included in the Euclidean path integral \cite{Fan:2021ntg,Heidenreich:2020pkc}. However, as we discussed in Sec.~\ref{sec:condensation}, the origin of the rotor is the Goldstone boson associated with the localized W-boson condensation of the monopole background. Thus, it seems that our winding rotor configuration and the particular 2d photon
configuration in \cite{Csaki:2024ajo} are not manifestly the same. We leave a detailed analysis of this problem to the future \cite{Csaki:2026toappear,Geng:2026toappear}.

Finally, let's comment on \cite{Brennan:2023tae}. One of the proposals in \cite{Brennan:2023tae} is a dynamical resolution of the energy conservation puzzle. The author started with the observation that if we start with a $\theta$-term in the 4d Lagrangian, the resulting rotor Lagrangian would also have an additional theta term \cite{Witten:1979ey}
\begin{equation}
    L_{\text{rotor}}=\frac{I}{2}\big(\partial_{t}\alpha(t)\big)^2+\frac{\theta}{2\pi}\partial_{t}\alpha(t)\,.
\end{equation}
Therefore, the rotor Hamiltonian is
\begin{equation}
    H_{\text{rotor}}=\frac{(\Pi_{\alpha}-\frac{\theta}{2\pi})^2}{2I}\,,\label{eq:rotorHtheta}
\end{equation}
where $\Pi_{\alpha}$ is the canonical momentum of the rotor. The author tried to show that the energy required to excite the rotor Hamiltonian Equ.~(\ref{eq:rotorHtheta}) wouldn't be large if a rotor vertex $e^{i\alpha(t)}$ is created in a physical process together with $\theta$ shifted by $2\pi$. The starting point of \cite{Brennan:2023tae} in our language is the tilded frame in Sec.~\ref{sec:thetachanging}. We provided a very simple proof of this effect. We worked out the canonical Hamiltonian in Equ.~(\ref{eq:Htheta}) and found that indeed there is an effective ``dynamical $\theta$" made of fermion fields. Therefore, a corroborated shift of ``dynamical $\theta$" with the rotor vertex in the scattering process guarantees that there wouldn't be any excitation with energy of order $I^{-1}$.

\section{Conclusions and Discussions}\label{sec:conclusions}

We have revisited massless fermion scattering off a minimally charged BPS ’t Hooft–Polyakov monopole, provided a new unifying picture to clarify the connections and resolve apparent conflicts between existing proposals. Starting from the monopole background and its electric collective coordinate, we derived the low-energy fermion–rotor description and examined its dynamics while retaining explicit fermion operators. Our analysis connects rotor dressing, twisted outgoing states, and instantonic configuration-induced anomalous correlators, showing how these mechanisms fit into a mutually consistent picture within the proposals’ common regime of applicability.

A central tool is a dual description obtained by adjoining a decoupled auxiliary scalar. This dual description permits an exact quantization and a direct calculation of the rotor vertex’s spectral function. Although the isolated rotor has a discrete excitation spectrum, the rotor vertex in the coupled system has spectral support extending to arbitrarily low energy in the infinite-volume limit. Physical charge transfer is governed by this continuous spectrum, resolving the apparent conflict between low-energy scattering and the excitation cost inferred from the isolated rotor. The alternative fermion in the titled frame provide a complementary canonical account of the same result through the correlated changes of rotor momentum and fermion charge.

The operator description also makes the nature of the outgoing excitation explicit. The fermion’s Heisenberg evolution produces a rotor dressing evaluated at the time of passage through the monopole core. Using the dual operator dictionary, we computed its commutation relation with the fermion field and recovered a twist action in the infrared limit. This establishes a direct connection between the dynamical rotor description and the interpretation of outgoing fermions as excitations in a twisted sector.

We further computed anomalous fermion correlators directly in the original fermion-rotor system. Winding configurations of the compact rotor generate normalizable fermion zero modes, whose saturation determines the allowed anomalous fermion insertions. Integrating over the remaining fermion and rotor modes gives the corresponding correlation functions without invoking cluster decomposition. This calculation makes the winding-sector origin of these processes explicit and connects it to their operator description. 

Several questions merit further study. A detailed matching should establish the relation between rotor winding and the topologically nontrivial configurations involving dynamical photons in the descriptions of monopole scattering in \cite{Csaki:2024ajo}. Extending the analysis in this paper to nonzero fermion masses, more general charge assignments of fermions, higher charged monopoles, and monopoles in other gauge groups would be interesting. Such extensions would sharpen the comparison between the existing proposals and broaden the range of monopole scattering processes accessible to our approach. We hope that, by unifying these different proposals, our work will help bridge communities with different expertise and lead toward a complete understanding of this longstanding puzzle in quantum field theory.

\section*{Acknowledgments}
We thank Diego García-Sepúlveda, Ziming Ji, Yikun Jiang, Shi Chen, Zhihan Liu, Vazha Loladze, Takemichi Okui,  Lisa Randall, Matt Reece, John Terning, Huy Tran, Huangyu Xiao, Juven Wang for helpful discussions. We thank Csaba Csáki, Yuta Hamada, Larissa Kiriliuk, Zhihan Liu, Rotem Ovadia, Lisa Randall, Matt Reece, John Terning, Huy Tran and  Huangyu Xiao for comments on the draft. We thank Chat-GPT and Claude for the assistance in clarifying the status of this problem and the relevant literature, the editing of the draft and the derivations in the Appendix. HG is supported by the Gravity, Spacetime, and Particle Physics (GRASP) Initiative from Harvard University.

\appendix

\section{Derivation for the Dual Description}\label{sec:derivation}
In this Appendix, we derive a dual description of the fermion-rotor system. This derivation starts with existing results in \cite{Polchinski:1984uw,Loladze:2025jsq}. For completeness and the convenience of use in the main text, we will repeat all their derivations. This also helps to set up some notations and conventions and we will fix a few minor errors in \cite{Loladze:2025jsq}.

The full action consists of Equ.~(\ref{eq:Polchinski}) and the rotor kinetic term\footnote{Note the difference between our conventions and the ones used in \cite{Loladze:2025jsq}. Our flavor number is $2N$ with theirs $N$ and our rotor is minus of their rotor.}
\begin{equation}
    S=\int dtdx\sum_{f=1}^{2N}\psi_{f}^{\dagger}(t,x)\Big(i\partial_{t}+i\partial_{x}+q_{f}\alpha(t)f'(x)\Big)\psi_{f}(t,x)+\int dt\frac{I}{2}\dot{\alpha}(t)^2-\int dt H_{\text{ct}}[\alpha]\,,
\end{equation}
where the counter term $H_{\text{ct}}[\alpha]$ will be fixed later. Let's define 
\begin{equation}
  \partial_{+}\equiv\partial_{t}+\partial_{x}\,, \quad D_f=\partial_{+}-iq_{f}\alpha(t)f'(x)\,. 
\end{equation}
For the sake of generality, we are interested in the generating functional\footnote{Note that this is not the most general generating functional as we haven't turned on a rotor source. We do this for convenience.}
\begin{equation}
\begin{split}
 Z[\xi,\xi^{\dagger}]&=\int  D\psi D\psi^{\dagger}D\alpha e^{iS+i\int dtdx\sum_{f=1}^{2N}(\xi_{f}^{\dagger}(t,x)\psi_{f}(t,x)+\psi^{\dagger}_{f}(t,x)\xi_{f}(t,x))}\,,\\&=\int D\alpha\prod_{f=1}^{2N}\det D_{f}e^{i\int dt\frac{I}{2}\dot{\alpha}(t)^2+\sum_{f=1}^{2N}\int dtdx\xi_{f}^{\dagger}D_{f}^{-1}\xi_{f}}\,.
 \end{split}
\end{equation}
Inside the above path integral, we can write
\begin{equation}
    D_{f}=e^{iq_{f}\lambda}\partial_{+}e^{-iq_{f}\lambda}=\partial_{+}(1-iq_{f}\lambda)\,,
\end{equation}
where 
\begin{equation}
    \lambda(t,x)=\partial_{+}^{-1}\alpha(t)f'(x)=\int dt'dx'G_{0}(t-t',x-x')\alpha(t')f'(x')\,,\label{eq:lambdaf}
\end{equation}
and $G_{0}(t-t',x-x')=\frac{1}{2\pi i}\frac{t-t'+x-x'}{(t-t')^2-(x-x')^2-i\epsilon^{+}}$ is the time-ordered two-point function of a free chiral fermion. The above variable $\lambda(t,x)$ can be expanded in momentum space as
\begin{equation}
    \lambda(t,x)=\int dx'\frac{d\omega}{2\pi}\frac{dk}{2\pi}e^{-i\omega t}\alpha(\omega)f'(x')\frac{-i(k+\omega)}{k^2-\omega^2-i\epsilon^{+}}e^{ik(x-x')}.
\end{equation}
In the IR limit, we can take $f'(x)=2\delta(x)$. Then, it is easy to see that \cite{Polchinski:1984uw,Loladze:2025jsq}
\begin{equation}
    \lambda(t,x)=2\int_{0}^{\infty}\frac{d\omega}{2\pi}\Big[\theta(x)\alpha(\omega)e^{-i\omega(t-x)}-\theta(-x)\alpha(-\omega)e^{i\omega(t-x)}\Big]\,.
\end{equation}
Hence, the generating functional can be written as
\begin{equation}
    \begin{split}
Z[\xi,\xi^{\dagger}]&=\text{ det}^{2N}\partial_{+}\int D\alpha \det(1-iq_{f}\lambda)e^{i\int dt\frac{I}{2}\dot{\alpha}(t)^2-i\int dt H_{ct}[\alpha]}\\&\times e^{-\sum_{f=1}^{2N}\int dtdxdt'dx'\xi_{f}^{\dagger}(t,x)e^{iq_{f}\lambda(t,x)}G_{0}(t-t',x-x')e^{-iq_{f}\lambda(t',x')}\xi_{f}(t',x')}\,,\label{eq:Zpartial}
    \end{split}
\end{equation}
To get the full generating functional, we have to re-exponentiate $\det(1-iq_{f}\lambda)$
\begin{equation}
    \det(1-iq_{f}\lambda)=e^{\Tr \log(1-iq_{f}\lambda)}\,,
\end{equation}
which can be computed diagrammatically by expanding the $\log$. We first note that as a functional operator we actually have
\begin{equation}
    \langle t, x|\lambda|t',x'\rangle=G_{0}(t-t',x-x')\alpha(t')f'(x')\,,
\end{equation}
and Equ.~(\ref{eq:lambdaf}) describes how the operator acts on standard operators.\footnote{For example, $\lambda\psi(t,x)= \int dt'dx'\langle t, x|\lambda|t',x'\rangle\psi(t',x')$ for which we have $\partial_{+}\lambda\psi(t,x)=\partial_{+}(\lambda(t,x))\psi(t,x)$.} It is easy to see that only one term survives.\footnote{This is easiest to see by induction. A general connected diagram from the expansion of $\Tr\log(1-iq_{f}\lambda)$ contains $n$ vertices, the free fermion propagators $G_{0ij}$ connecting these vertices, and the integral over the possible positions of these $n$ vertices. Therefore, all of these vertices are symmetrized, and the diagram can be thought of as a circle with $n$ points. The result of a diagram with $n$ vertices can be inductively obtained from the result of the diagram with $n-1$ vertices by inserting one vertex into the $(n-1)-$pointed circle and summing over all possible such insertions. The insertion between $i$ and $j$ replaces $G_{0ij}$ by $G_{0in}G_{0nj}=G_{0ij}(G_{0in}-G_{0jn})$, i.e. producing a factor $(G_{0in}-G_{0jn})$. Thus, summing over the possible insertions gives a zero result. } The result is
\begin{equation}
    \begin{split}
 \det(1-iq_{f}\lambda)&=e^{\Tr \log(1-iq_{f}\lambda)}=e^{\sum_{f=1}^{2N}q_{f}^2\frac{1}{2}\int dt'dx' dtdx\langle t, x|\lambda|t',x'\rangle\langle t', x'|\lambda|t,x\rangle}\,,\\&=e^{-\sum_{f=1}^{2N}q_{f}^2\frac{1}{2}\int dt'dx' dtdx \alpha(t)f'(x)G_{0}(t-t',x-x')^{2}\alpha(t')f'(x')}\,,\\&=e^{-\sum_{f=1}^{2N}q_{f}^2\frac{4}{2}\int dt' dt \alpha(t)G_{0}(t-t',0)^{2}\alpha(t')}\,.\label{eq:det}
    \end{split}
\end{equation}
The exponent in Equ.~(\ref{eq:det}) can be computed in the mixed momentum and real space as 
\begin{equation}
    \begin{split}
\frac{1}{2}\int dt' dt \alpha(t)G_{0}(t-t',0)^{2}\alpha(t')&=-\frac{1}{8\pi^2}\int dt'dt\frac{d\omega'}{2\pi}\frac{d\omega}{2\pi}\alpha(\omega)\alpha(\omega')e^{-i\omega t-i\omega' t'}\frac{(t-t')^2}{\big((t-t')^2-i\epsilon^{+}\big)^2}\,,\\&=-\frac{1}{8\pi^2}\int d\tau\frac{d\omega}{2\pi}\alpha(\omega)\alpha(-\omega)e^{-i\omega\tau}\frac{\tau^2}{(\tau^2-i\epsilon^{+})^2}\,,\\&\equiv-\frac{1}{8\pi^2}\int \frac{d\omega}{2\pi}\alpha(\omega)\alpha(-\omega)K_{\epsilon}(\omega)\,,\label{eq:epsilonreg}
    \end{split}
\end{equation}
where we have defined
\begin{equation}
    K_{\epsilon}(\omega)=\int_{-\infty}^{\infty} d\tau\frac{\tau^2}{(\tau^2-i\epsilon^{+})^2}e^{-i\omega\tau}\,.
\end{equation}
This quantity is divergent as $\epsilon^{+}\rightarrow0$ and independent of $\omega$. We will follow \cite{Polchinski:1984uw} to consider the regulated version
\begin{equation}
\begin{split}
    K^{\text{reg}}(\omega)&=\lim_{\epsilon^{+}\rightarrow0}\Big(K_{\epsilon}(\omega)-K_{\epsilon}(0)\Big)\,,\\&=\lim_{\epsilon^{+}\rightarrow0}\int_{-\infty}^{\infty}d\tau \frac{\tau^2}{(\tau^2-i\epsilon^{+})^2}(\cos\omega\tau-1)\,,\\&=\lim_{\epsilon^{+}\rightarrow0}\int_{-\infty}^{\infty}d\tau\frac{\cos\omega\tau-1}{(\tau^2-i\epsilon^{+})}\,,\\&=-\lim_{\epsilon^{+}\rightarrow0}\omega\int_{-\infty}^{\infty}d\tau\frac{\sin\omega\tau}{\tau-i\epsilon^{+}\text{sign}{\tau}}\,,\\&=-\omega\lim_{\epsilon^{+}\rightarrow0}\Big(\int_{-\infty}^{-\epsilon^{+}}d\tau\frac{\sin\omega\tau}{\tau}+\int_{\epsilon^{+}}^{\infty}d\tau\frac{\sin\omega\tau}{\tau}\Big)\,,\\&=-\omega\text{Im}\lim_{\epsilon^{+}\rightarrow0}\Big(\int_{-\infty}^{-\epsilon^{+}}d\tau\frac{e^{i\omega\tau}}{\tau}+\int_{\epsilon^{+}}^{\infty}d\tau\frac{e^{i\omega\tau}}{\tau}\Big)\,,\\&=-\pi\abs{\omega}\,.\label{eq:ctfixed}
    \end{split}
\end{equation}
As a result, we have the regulated determinant, i.e. the quantum induced rotor action, as
\begin{equation}
    \det(1-iq_{f}\lambda)=e^{-\sum_{f=1}^{2N}\frac{q_{f}^{2}}{2\pi}\int_{-\infty}^{\infty} \frac{d\omega}{2\pi}\abs{\omega}\alpha(\omega)\alpha(-\omega)+i\int dt H_{\text{ct}}[\alpha]}\,,\quad H_{\text{ct}}[\alpha]=-i\sum_{f=1}^{2N}\frac{q_{f}^2}{2\pi^2}K_{\epsilon}(0)\alpha(t)^2
\end{equation}
for which the regulated term is shift symmetric if one introduces an IR regulator first. Therefore, the generating functional Equ.~(\ref{eq:Zpartial}) can be written as
\begin{equation}
    \begin{split}
Z[\xi,\xi^{\dagger}]&=\text{ det}^{2N}\partial_{+}\int D\alpha e^{i\int dt\frac{I}{2}\dot{\alpha}(t)^2-\sum_{f=1}^{2N}\frac{q_{f}^{2}}{2\pi}\int_{-\infty}^{\infty} \frac{d\omega}{2\pi}\abs{\omega}\alpha(\omega)\alpha(-\omega)}\\&\times e^{-\sum_{f=1}^{2N}\int dtdxdt'dx'\xi_{f}^{\dagger}(t,x)e^{iq_{f}\lambda(t,x)}G_{0}(t-t',x-x')e^{-iq_{f}\lambda(t',x')}\xi_{f}(t',x')}\,.\label{eq:Z}
\end{split}
\end{equation}
This is useful for us in Sec.~\ref{sec:correlators} to analyze the correlators. For later convenience, let's regulate the counter term in momentum space. To obtain this momentum regulated counter term, let's compute the exponent of Equ.~(\ref{eq:det}) by expanding first the propagators in the momentum space
\begin{equation}
\begin{split}
   &\frac{1}{2}\int dtdt'\alpha(t)G^{2}(t-t',0)\alpha(t')= \frac{1}{2}\int dtdt'\frac{d\omega}{2\pi}\frac{dk}{2\pi}\frac{d\omega'}{2\pi}\frac{dk'}{2\pi} \alpha(t)\alpha(t')e^{-i\omega(t-t')}e^{-i\omega'(t-t')}\\&\quad\quad\quad\quad\quad\quad\quad\quad\quad\quad\times\frac{-1}{(\omega-k+i0\text{sign}(k))(\omega'-k'+i0\text{sign}(k'))}\,,\\&= \frac{1}{2}\int\frac{d\omega}{2\pi}\frac{dk}{2\pi}\frac{d\omega'}{2\pi}\frac{dk'}{2\pi}\frac{-\alpha(-\omega-\omega')\alpha(\omega+\omega')}{(\omega-k+i0\text{sign}(k))(\omega'-k'+i0\text{sign}(k'))}\,,\\&=- \frac{1}{2}\int\frac{d\nu}{2\pi}\frac{d\omega'}{2\pi}\frac{dk'}{2\pi}\frac{dk}{2\pi}\frac{\alpha(\nu)\alpha(-\nu)}{(\nu-\omega'-k+i0\text{sign}(k))(\omega'-k'+i0\text{sign}(k'))}\,,\\&=i \frac{1}{2}\int\frac{d\nu}{2\pi}\int_{0}^{\infty}\frac{dk}{2\pi}\int_{0}^{\infty}\frac{dk'}{2\pi}\alpha(\nu)\alpha(-\nu)\Big[\frac{1}{\nu-k-k'+i\epsilon^{+}}-\frac{1}{\nu+k+k'-i\epsilon^{+}}\Big]\,,\\&=i \frac{1}{2}\int\frac{d\nu}{2\pi}\int_{0}^{\infty}\frac{dp}{2\pi}\int_{0}^{p}\frac{dk'}{2\pi}\alpha(\nu)\alpha(-\nu)\Big[\frac{1}{\nu-p+i\epsilon^{+}}-\frac{1}{\nu+p-i\epsilon^{+}}\Big]\,.
   \end{split}
\end{equation}
We can regulate the $p$ integral by introducing a UV cutoff $\Lambda$. Thus, we have
\begin{equation}
    \frac{1}{2}\int dtdt'\alpha(t)G^{2}(t-t',0)\alpha(t')=i \frac{1}{2}\int\frac{d\nu}{2\pi}\int_{0}^{\Lambda}\frac{dp}{2\pi}\int_{0}^{p}\frac{dk'}{2\pi}\alpha(\nu)\alpha(-\nu)\Big[\frac{1}{\nu-p+i\epsilon^{+}}-\frac{1}{\nu+p-i\epsilon^{+}}\Big]\,.
\end{equation}
Comparing with Equ.~(\ref{eq:epsilonreg}), we have
\begin{equation}
    K_{\epsilon}(\omega)=K_{\Lambda}(\omega)=-i\int_{0}^{\Lambda}dp\frac{2p^2}{\omega^2-p^2+i\epsilon^{+}}\,.
\end{equation}
Thus, we have
\begin{equation}
    K_{\epsilon}(0)=K_{\Lambda}(0)=2i\Lambda\,,
\end{equation}
and with the momentum space cutoff the counter term can be written as
\begin{equation}
    H_{\text{ct}}[\alpha]=\sum_{f=1}^{2N}\frac{q_{f}^2}{\pi^2}\alpha(t)^2 \Lambda\,.\label{eq:momentumCT}
\end{equation}
In fact, if one uses the heat kernel method to compute the fermion determinant there is no need to add such a counter term (see Appendix.~\ref{sec:normalization}).

\subsection{The Dual Path Integral}
For the purpose of deriving the dual description in Sec.~\ref{sec:dual}, let's consider the source free partition function
\begin{equation}
    Z=\text{ det}^{2N}\partial_{+}\int D\alpha e^{i\int dt\frac{I}{2}\dot{\alpha}(t)^2-\sum_{f=1}^{2N}\frac{q_{f}^{2}}{2\pi}\int_{-\infty}^{\infty} \frac{d\omega}{2\pi}\abs{\omega}\alpha(\omega)\alpha(-\omega)}\,.
\end{equation}
This partition function can be equivalently written as
\begin{equation}
    \begin{split}
Z&=\frac{1}{Z^{D}_{\text{auxiliary}}}\int D\Psi_{f}D\Psi^{\dagger}_{f}D\eta D\phi D\alpha e^{i\int dtdx\sum_{f}\Psi_{f}^{\dagger}(t,x)\big(i\partial_{t}+i\partial_{x}\big)\Psi_{f}(t,x)}\\\times &e^{i\int dt\frac{I}{2}\dot{\alpha}(t)^2-i\int dt\int_{0}^{\infty}dx\frac{\sum_{f=1}^{2N}q_{f}^2}{8\pi}\partial_{\mu}\phi(t,x)\partial^{\mu}\phi(t,x)+i\int dt\eta (\phi(t,0)-2\alpha(t))}\,,
    \end{split}
\end{equation}
where the auxiliary partition function is
\begin{equation}
    Z^{D}_{\text{auxiliary}}=\int D\Phi e^{-i\int dt\int_{0}^{\infty}dx\frac{\sum_{f=1}^{2N}q_{f}^2}{8\pi}\partial_{\mu}\Phi(t,x)\partial^{\mu}\Phi(t,x)}\,,
\end{equation}
with $\Phi(t,x)$ obeying the Dirichlet boundary condition at $x=0$ which explains the superscript $D$. 

In summary, adding an auxiliary boson on the positive half plane $x>0$ obeying the Dirichlet boundary condition at x=0 and a counter term Equ.~(\ref{eq:momentumCT}), we obtained a dual description of the fermion-rotor-auxiliary boson theory
\begin{equation}
\begin{split}
    ZZ^{D}_{\text{auxiliary}}&=\int D\Psi_{f}D\Psi^{\dagger}_{f}D\eta D\phi D\alpha e^{i\int dtdx\sum_{f}\Psi_{f}^{\dagger}(t,x)\big(i\partial_{t}+i\partial_{x}\big)\Psi_{f}(t,x)}\\\times &e^{i\int dt\frac{I}{2}\dot{\alpha}(t)^2-i\int dt\int_{0}^{\infty}dx\frac{\sum_{f=1}^{2N}q_{f}^2}{8\pi}\partial_{\mu}\phi(t,x)\partial^{\mu}\phi(t,x)+i\int dt\eta (\phi(t,0)-2\alpha(t))}\,.\label{eq:partitionfucntionduality}
    \end{split}
\end{equation}
This dual description consists of free fermions $\Psi_{f}(t,x)$, the rotor $\alpha(t)$, and the half-plane boson $\phi(t,x)$ whose boundary profile at $x=0$ is constrained to be that of the rotor. 

\subsection{The Operator Dictionary}\label{sec:operatordic}
However, to make use of this dual description, we have to provide the canonical operator dictionary between this dual description and the original fermion-rotor-auxiliary boson description. This dictionary can be motivated by the observation that on both sides of Equ.~(\ref{eq:partitionfucntionduality}), we have a free sector decoupled from the rotor and a sector that is coupled to the rotor. However, the free sector in the original description contains fewer degrees of freedom than the free sector in the dual description, while the interacting sector in the dual description contains fewer degrees of freedom than the interacting sector in the original description. Hence, the free sector in the original description should map into the interacting sector in the dual description, and the interacting sector in the dual description should map into the free sector in the original description. With this principle understood, we can work out the dictionary piece by piece and check the consistency at the end.

Let's list the results and provide a consistency check at the end. First, we introduce the following notations
\begin{equation}
    Q=\sum_{f=1}^{2N}q_{f}^2\,,\quad \gamma=\frac{Q}{4\pi}\,,\quad j_{R}(t,x)=\sum_{f=1}^{2N}q_{f}:\Psi^{\dagger}_{f}\Psi_{f}:(t,x)=\int \frac{dk}{2\pi} e^{ikx}j_{R}(t,k)\,,
\end{equation}
where $::$ denotes the normal ordering. For convenience, we will unfold the half-plane bosons for both the dual description and the original description. We introduce the unfolded boson current
\begin{equation}
    B_{\phi,R}(t,x)=\gamma\phi'(t,|x|)-\text{sign}(x)\gamma\partial_{t}\phi(t,|x|)=\int \frac{dk}{2\pi}e^{ikx}B_{\phi,R}(t,k)
\end{equation}
where 
\begin{equation}
    \phi'(t,|x|)=\frac{d \phi(t,r)}{dr}\Big|_{r=\abs{x}}\,.
\end{equation}
Let's denote the scalar Dirichlet modes
\begin{equation}
   \phi_{k}(t)=\sqrt{\frac{2}{\pi}}\int_{0}^{\infty} dr \phi(t,r)\sin kr\,,\quad \Pi_{k}(t)=\gamma\partial_{t} \phi_{k}(t)\,,
\end{equation}
in terms of which we have
\begin{equation}
    B_{\phi,R}(t,k)=-4\gamma\alpha(t)+\sqrt{2\pi}\Big(\gamma k\phi_{k}(t)+i\Pi_{k}(t)\Big)\,,
\end{equation}
where we used the boundary condition
\begin{equation}
    \phi(t,0)=2\alpha(t)\,.
\end{equation}
The canonical boson current is defined as
\begin{equation}
    B_{R}^{\text{can}}(t,x)=B_{\phi,R}(t,x)+4\gamma\alpha(t)\delta(x)=\int \frac{dk}{2\pi}e^{ikx}B_{R}^{\text{can}}(t,k)\,,
\end{equation}
where
\begin{equation}
    B_{R}^{\text{can}}(t,k)=\sqrt{2\pi}\Big(\gamma k\phi_{k}(t)+i\Pi_{k}(t)\Big)\,.
\end{equation}
To provide the operator dictionary, let's further introduce
\begin{equation}
    \Delta(t,k)=j_{R}(t,k)-B_{R}^{\text{can}}(t,k)\,,\quad N_{\Lambda}=\frac{1}{2Q}\int_{0}^{\Lambda}\frac{dk}{k}\Delta(t,-k)\Delta(t,k)\,.
\end{equation}
The operator dictionary is given by the unitary operator
\begin{equation}
    U_{\Lambda}=e^{i\pi N_{\Lambda}}\,.
\end{equation}
Then the transformation of all the operators can be worked out using the following canonical commutation relations
\begin{equation}
\begin{split}
    &[j_{R}(t,k),\Psi_{f}(t,x)]=-q_{f}e^{-ikx}\Psi_{f}(t,x)\,,\quad[j_{R}(t,k),j_{R}(t,k')]=Qk\delta(k+k')\,,\\&[B_{R}^{\text{can}}(t,k),B_{R}^{\text{can}}(t,k')]=Qk\delta(k+k')\,,\quad [\phi_{k}(t),\Pi_{k'}(t)]=i\delta(k-k')\,,
    \end{split}
\end{equation}
together with the fermion's canonical commutation relation, which we do not write out here. The operator dictionary is then
\begin{equation}
\begin{split}
    &j_{L}(t,x)\equiv\sum_{f=1}^{2N}q_{f}:\psi^{\dagger}_{f}\psi_{f}:(t,x)=U_{\Lambda}j_{R}(t,x)U^{\dagger}_{\Lambda}=B_{R}^{\text{can}}(t,x)\,,\\&\rho_{L,f}(t,x)\equiv:\psi^{\dagger}_{f}\psi_{f}:(t,x)=U_{\Lambda}\rho_{R,f}(t,x)U^{\dagger}_{\Lambda}=\rho_{R,f}(t,x)+\frac{q_f}{Q}\Big[B_{R}^{\text{can}}(t,x)-j_{R}(t,x)\Big]\,,\\   &\psi_{f}(t,x)=U_{\Lambda}\Psi_{f}(t,x)U^{\dagger}_{\Lambda}=e^{A_{f,R}}\Psi_{f}e^{-A_{f,R}^{\dagger}}(t,x)\,,\\&B_{L}(t,x)\equiv\gamma\Phi'(t,|x|)-\text{sign}(x)\gamma\partial_{t}\Phi(t,|x|)=U_{\Lambda}B_{R}^{\text{can}}(t,x)U^{\dagger}_{\Lambda}=j_{R}(t,x)\,,\\&\Phi(t,r)=U_{\Lambda}\phi^{\text{sin}}_{\Lambda}(t,r)U^{\dagger}_{\Lambda}=\frac{1}{2\gamma}\int_{-r}^{r}dx j_{R}(t,x)\,,\\&\Pi_{\Phi}(t,r)\equiv\gamma\partial_{t}\Phi(t,x)=U_{\Lambda}\Pi_{\phi}(t,x)U^{\dagger}_{\Lambda}=\frac{1}{2}(j_{R}(t,-r)-j_{R}(t,r))\,,\\&\alpha(t)=U_{\Lambda}\alpha(t)U^{\dagger}_{\Lambda}=\alpha(t)\,,\label{eq:dictionary}
    \end{split}
\end{equation}
where we defined
\begin{equation}
    A_{f,R}(t,x)=\frac{q_{f}}{Q}\int_{0}^{\Lambda}\frac{dk}{k}e^{-ikx}\Big(j_{R}(t,-k)-B_{R}^{\text{can}}(t,-k)\Big)\,,
\end{equation}
and
\begin{equation}
    \phi^{\text{sin}}_{\Lambda}(t,r)=\sqrt{\frac{2}{\pi}}\int_{0}^{\Lambda}dk\phi_{k}(t)\sin kr\,.
\end{equation}
We have explicitly introduced the cutoff $\Lambda$ for convenience in dealing with the counter term. As a consistency check of the operator dictionary Equ.~(\ref{eq:dictionary}), we can see how the Hamiltonian of the two descriptions are mapped to each other. The Hamiltonian of the original theory with the counter term is
\begin{equation}
    H_{L}=H_{\psi,0}+H_{\Phi,0}+\frac{p_{\alpha}(t)^2}{2I}-2\alpha(t)j_{L}(t,0)+\frac{4\gamma\Lambda}{\pi}
\alpha(t)^2\,,
\end{equation}
where we have
\begin{equation}
\begin{split}
    H_{\psi,0}&=-i\sum_{f=1}^{2N}\int dx :\psi^{\dagger}_{f}\partial_{x}\psi_{f}:(t,x)\,,\\&=\frac{1}{4\gamma}\int dx:j_{L}(t,x)^2:+\pi\int dx\sum_{f=1}^{2N}:(\rho_{f}(t,x)-\frac{q_{f}}{Q}j_{L}(t,x))^2:\,,
    \end{split}
\end{equation}
and
\begin{equation}
    H_{\Phi,0}=\int_{0}^{\infty} dr\Big[\frac{\Pi_{\Phi}^2}{2\gamma}+\frac{\gamma}{2}(\partial_{r}\Phi)^2\Big]=\frac{1}{4\gamma}\int dx:B_{L}(t,x)^2:\,.
\end{equation}
Using the operator dictionary Equ.~(\ref{eq:dictionary}), we have
\begin{equation}
\begin{split}
    H_{\psi,0}&=\frac{1}{4\gamma}\int dx :B_{R}^{\text{can}}(t,x)^2:+\pi\int dx\sum_{f=1}^{2N}:(\rho_{R,f}(t,x)-\frac{q_{f}}{Q}j_{R}(t,x))^2:\,,\\
    H_{\Phi,0}&=\frac{1}{4\gamma}\int dx:j_{R}(t,x)^{2}:\,,\\
    \alpha(t)j_{L}(t,0)&=\alpha(t)B_{R}^{\text{can}}(t,0)\,.\label{eq:HHH}
    \end{split}
\end{equation}
Combining Equ.~(\ref{eq:HHH}) with the rotor kinetic term and the counter term, we have
\begin{equation}
\begin{split}
    H_{L}&=H_{\Psi,0}+\frac{p_{\alpha}(t)^2}{2I}+\frac{1}{4\gamma}\int dx :B_{R}^{\text{can}}(t,x)^2:-2\alpha(t)B_{R}^{\text{can}}(t,0)+\frac{4\gamma\Lambda}{\pi}\alpha(t)^2\,,\\&=H_{\Psi,0}+\frac{p_{\alpha}(t)^2}{2I}+\frac{1}{4\gamma}\int dx :\Big(B_{R}^{\text{can}}(t,x)-4\gamma\alpha(t)\delta(x)\Big)^2:\,,\\&=H_{\Psi,0}+\frac{p_{\alpha}(t)^2}{2I}+\int_{0}^{\infty} dr\Big[\frac{\gamma}{2}(\partial_{r}\phi(t,r))^2+\frac{\Pi_{\phi}(t,r)^2}{2\gamma}\Big]\,,\\&=H_{R}\,,
    \end{split}
\end{equation}
where we used the regulated $\delta(x)^2$
\begin{equation}
   \int dx \delta(x)^2=\delta(0)=\int_{-\Lambda}^{\Lambda} \frac{dk}{2\pi}=\frac{\Lambda}{\pi}\,.
\end{equation}
Thus, we can see that the Hamiltonians of the original theory and the dual theory are the same. This result will be important for us, as the dual theory is much easier to deal with, and we can understand the dynamics of the operators in the original theory using the dual theory.

\section{Fixing the Constant Factor in the Path Integral}\label{sec:normalization}
In this Appendix,  we explain how the factor $\mathcal{N}(\beta)$ in Equ.~(\ref{eq:2pt}) and Equ.~(\ref{eq:4pt}) is fixed. We will consider the case with a general winding number $w<0$ with $q_{f}=\frac{1}{2}$. The same method easily generalizes to $w>0$.

First, we note that the zero mode functions in Equ.~(\ref{eq:zeromodes}) are not normalized. More precisely, we have
\begin{equation}
    \psi^{0}_{f,r}(z)=e^{\frac{i}{2}f(x)\alpha_{0}+\frac{i}{2}(f(x)-1)\sum_{n>0}\alpha_{n}z^{n}+\frac{i}{2}(f(x)+1)\sum_{n<0}\alpha_{n}z^{n}}z^{r+\frac{w}{2}f(x)}\,,\label{eq:rzeromode}
\end{equation}
and the Gram matrix of the inner products between two functions is
\begin{equation}
    G^{f}_{rs}[\alpha]\equiv(\psi^{0}_{f,r},\psi^{0}_{f,s})=\int_{0}^{\beta}d\tau\int dx\psi^{0*}_{f,r}(z)\psi^{0}_{f,s}(z)\,.
\end{equation}
Thus, for each flavor we have an additional factor from the fermion path integral\footnote{We note that in principle we also need a factor $e^{-S_{\text{ct}}[\tilde{\alpha}]}$, where $\tilde{\alpha}$ is the periodic part of the rotor. However, here we will use the heat kernel method to compute the determinant ratio for which there is no divergence like Equ.~(\ref{eq:momentumCT}). Thus, we will ignore the counter term in this section.}
\begin{equation}
    \frac{e^{\sum_{n>0}\frac{n}{2}\alpha_{n}\alpha_{-n}}\det'\mathcal{D}_{\alpha}}{\det \mathcal{D}_{0}\sqrt{\det G^{f}[\alpha]}}\,,
\end{equation}
where
\begin{equation}
    \mathcal{D}_{\alpha}=i\partial_{\tau}+\partial_{x}-\frac{i}{2}\alpha(\tau)f'(x)\,,\quad \mathcal{D}_{0}=i\partial_{\tau}+\partial_{x}\,.
\end{equation}
Let's first consider
\begin{equation}
    C_{w}[\alpha]\equiv\Big|\frac{\det'\mathcal{D}_{\alpha}}{\det \mathcal{D}_{0}}\Big|\,.\label{eq:targetC}
\end{equation}
To compute Equ.~(\ref{eq:targetC}), let's define
\begin{equation}
    \alpha_{s}(\tau)=\alpha_{0}+\frac{2\pi w\tau}{\beta}+s\tilde{\alpha}(\tau)\,,\quad 0\leq s\leq1\,,
\end{equation}
where we remember that $\tilde{\alpha}(\tau)$ is the periodic part of the rotor. We denote $C_{w}[\alpha_{s}]$ as $C_s$ and $\mathcal{D}_{\alpha_{s}}$ as $\mathcal{D}_{(s)}$. We further define
\begin{equation}
    T_{s}=e^{\frac{i}{2}(f(x)-1)s\sum_{n>0}\alpha_{n}z^{n}+\frac{i}{2}(f(x)+1)s\sum_{n<0}\alpha_{n}z^{n}}\,,\quad A=\dot{T}_{s}T^{-1}_{s}=(\frac{d}{ds}T_{s})T_{s}^{-1}\,,
\end{equation}
which implies
\begin{equation}
    \mathcal{D}_{(s)}=T_{s}\mathcal{D}_{(0)}T_{s}^{-1}\,,\quad \dot{\mathcal{D}}_{s}=[A,\mathcal{D}_{(s)}].
\end{equation}

Furthermore, we have
\begin{equation}
\begin{split}
\frac{d}{ds}\log C_{s}&=\text{Re}\Tr'\mathcal{D}_{(s)}^{-1}\dot{\mathcal{D}}_{s}=\text{Re}\Tr'\mathcal{D}_{(s)}^{-1}A\mathcal{D}_{(s)}-\text{Re}\Tr'\mathcal{D}_{(s)}^{-1}\mathcal{D}_{(s)}A\,,\\&=\text{Re}\Tr\mathcal{D}_{(s)}^{-1}A\mathcal{D}_{(s)}-\text{Re}\Tr'\mathcal{D}_{(s)}^{-1}\mathcal{D}_{(s)}A\,,\\&=\text{Re}\Tr\mathcal{D}_{(s)}\mathcal{D}_{(s)}^{-1}A-\text{Re}\Tr'\mathcal{D}_{(s)}^{-1}\mathcal{D}_{(s)}A\,,\\&=\text{Re}\Tr\Big(\mathcal{D}_{(s)}\mathcal{D}_{(s)}^{-1}A-\mathcal{D}_{(s)}^{-1}\mathcal{D}_{(s)}A\Big)+\text{Re}\Tr P_{s}A\,,\label{eq:Cdifferential}
\end{split}
\end{equation}
where $P_{s}$ is the projection to the zero mode space of $\mathcal{D}_{(s)}$ and $\mathcal{D}^{-1}_{s}$ in the inverse of $D_{(s)}$ in the domain with the zero modes projected out. Let's introduce the vector $V_{s}=T_{s}V_{0}$ whose $r$-th component is the $r$-th unnormalized zero mode, i.e. for $V_{0}$ it is Equ.~(\ref{eq:rzeromode}). We have\footnote{As a consistency check, it is easy to see $P_{s}P_{s}=P_{s}$.}
\begin{equation}
    G_{s}=V_{s}^{\dagger}V_{s}\,,\quad P_{s}=V_{s}G_{s}^{-1}V^{\dagger}_{s}\,.
\end{equation}
Thus, we have
\begin{equation}
    \dot{G}_{s}=V_{s}^{\dagger}(A^{\dagger}+A)V_{s}\,,
\end{equation}
and so
\begin{equation}
    \text{Re}\Tr P_{s}A=\Tr\frac{P_{s}A+A^{\dagger}P_{s}^{\dagger}}{2}=\Tr\frac{G_{s}^{-1}\dot{G}_{s}}{2}=\frac{1}{2}\frac{d}{ds}\log\det G_{s}\,.
\end{equation}
The first term in the final expression of Equ.~(\ref{eq:Cdifferential}) can be computed using the heat kernel method. For $\mathcal{D}_{(s)}$, we have the adjoint operator
\begin{equation}
    \mathcal{D}_{(s)}^{\dagger}=i\partial_{\tau}-\partial_{x}+\frac{i}{2}\alpha_{s}(\tau)f'(x)\,,
\end{equation}
which has no normalizable zero modes when $\mathcal{D}_{(s)}$ does. Thus, we can do the following regularization
\begin{equation}
    \begin{split}
\text{Re}\Tr\Big(\mathcal{D}_{(s)}\mathcal{D}_{(s)}^{-1}A-\mathcal{D}_{(s)}^{-1}\mathcal{D}_{(s)}A\Big)&=\lim_{\epsilon\rightarrow0}\text{Re}\Tr\Big((e^{-\epsilon \mathcal{D}_{(s)}\mathcal{D}^{\dagger}_{(s)}}-e^{-\epsilon \mathcal{D}^{\dagger}_{(s)}\mathcal{D}_{(s)}})A\Big)\,,\\&=\int dxd\tau A(\tau,x)(e^{-\epsilon \mathcal{D}_{(s)}\mathcal{D}^{\dagger}_{(s)}}-e^{-\epsilon \mathcal{D}^{\dagger}_{(s)}\mathcal{D}_{(s)}})(\tau,x;\tau,x)\,,
    \end{split}
\end{equation}
for which we have
\begin{equation}
    \begin{split}
        \mathcal{D}_{(s)}\mathcal{D}^{\dagger}_{(s)}&=-\partial_{x}^{2}-\partial_{\tau}^{2}-\frac{1}{2}\alpha_{s}'(\tau)f'(x)+\frac{1}{4}\alpha_{s}(\tau)^{2} f'(x)^2+\frac{i}{2}\alpha_{s}(\tau)(\partial_{x}f'(x)+f'(x)\partial_{x})\,,\\
        \mathcal{D}^{\dagger}_{(s)}\mathcal{D}_{(s)}&=-\partial_{x}^{2}-\partial_{\tau}^{2}+\frac{1}{2}\alpha_{s}'(\tau)f'(x)+\frac{1}{4}\alpha_{s}(\tau)^{2} f'(x)^2+\frac{i}{2}\alpha_{s}(\tau)(\partial_{x}f'(x)+f'(x)\partial_{x})\,.
    \end{split}
\end{equation}
Using the short distance expansion of the heat kernel, we have
\begin{equation}
    (e^{-\epsilon \mathcal{D}_{(s)}\mathcal{D}^{\dagger}_{(s)}}-e^{-\epsilon \mathcal{D}^{\dagger}_{(s)}\mathcal{D}_{(s)}})(\tau,x;\tau,x)=\frac{1}{4\pi\epsilon}(\epsilon\alpha_{s}'(\tau)f'(x))=\frac{\alpha_{s}'(\tau)f'(x)}{4\pi}\,.
\end{equation}
This implies
\begin{equation}
    \text{Re}\Tr\Big(\mathcal{D}_{(s)}\mathcal{D}_{(s)}^{-1}A-\mathcal{D}_{(s)}^{-1}\mathcal{D}_{(s)}A\Big)=\frac{1}{2\pi}\int_{0}^{\beta}d\tau A(\tau,0)\alpha_{s}'(\tau)=-s\sum_{n>0}n\alpha_{n}\alpha_{-n}\,.
\end{equation}
In summary, we have
\begin{equation}
    \frac{d}{ds}\log C_{s}=-s\sum_{n>0}n\alpha_{n}\alpha_{-n}+\frac{1}{2}\frac{d}{ds}\log\det G_{s}\,.
\end{equation}
Integrating $s$ from $0$ to $1$, we get
\begin{equation}
    C_{w}[\alpha]=e^{-\frac{1}{2}\sum_{n>0}n\alpha_{n}\alpha_{-n}}\sqrt{\frac{\det G^{f}[\alpha]}{\det G^{f}[\alpha-\tilde{\alpha}]}}C_{w}[\alpha-\tilde{\alpha}]\,.
\end{equation}
Putting everything together, we have
\begin{equation}
   \Bigg| \frac{e^{\sum_{n>0}\frac{n}{2}\alpha_{n}\alpha_{-n}}\det'\mathcal{D}_{\alpha}}{\det \mathcal{D}_{0}\sqrt{\det G^{f}[\alpha]}}\Bigg| =\Bigg| \frac{\det'\mathcal{D}_{\alpha-\tilde{\alpha}}}{\det \mathcal{D}_{0}\sqrt{\det G^{f}[\alpha-\tilde{\alpha}]}}\Bigg| \,.\label{eq:simpliedtarget}
\end{equation}

Second, let's compute the simplified quantity Equ.~(\ref{eq:simpliedtarget}). 
\begin{equation}
    \frac{\det'(i\partial_{\tau}+\partial_{x}-\frac{i}{2}\alpha_{w}(\tau)f'(x))}{\det(i\partial_{\tau}+\partial_{x}-\frac{i}{2}\alpha_{w=0}(\tau)f'(x))}\,,
\end{equation}
for each fermion flavor with $\alpha_{w}(\tau)=\frac{2\pi\tau w}{\beta}$.\footnote{The constant mode $\alpha_{0}$ can be removed by a unitary transform $U=e^{\frac{i}{2}\alpha_{0}f(x)}$ which doesn't change the result of the determinant. However, the winding mode $\alpha_{w}(\tau)$ cannot be removed by a unitary transform. Thus, the result of the determinant is independent of the zero mode but depends on the winding mode.} The ratio between the two determinants can be computed if we first IR regulate the spatial direction using a box of size $L$ and Dirichlet boundary conditions at the ends $x=\pm\frac{L}{2}$. Thus, we have
\begin{equation}
\begin{split}
\frac{\det'(\mathcal{D}_{\alpha-\tilde{\alpha}})}{\det(\mathcal{D}_{0})}&=\frac{\det'(i\partial_{\tau}+\partial_{x}-i\frac{2\pi w\tau}{\beta}\delta(x))}{\det(i\partial_{\tau}+\partial_{x})}\equiv\frac{\det'\mathcal{D}_{w}}{\det\mathcal{D}_{0}}\,.\label{eq:targetratio}
\end{split}
\end{equation}
To proceed, we can reduce the 2d differential operator in the numerator to a family of 1d differential operators by considering
\begin{equation}
    e^{\frac{2\pi i(n+\frac{1}{2})\tau}{\beta}}e^{i\frac{2\pi w\tau}{\beta}\Theta(x)}h_{n}(x)\,,
\end{equation}
Thus, the 2d differential operator becomes the following family of 1d differential operators
\begin{equation}
    \mathcal{D}_{n}=-\frac{2\pi(n+\frac{1}{2})}{\beta}-\frac{2\pi w}{\beta}\Theta(x)+\partial_{x}\,,
\end{equation}
which have adjoints
\begin{equation}
    \mathcal{D}_{n}^{\dagger}=-\frac{2\pi(n+\frac{1}{2})}{\beta}-\frac{2\pi w}{\beta}\Theta(x)-\partial_{x}\,.
\end{equation}
The determinants can be understood by considering
\begin{equation}
    H_{n}\equiv\mathcal{D}_{n}^{\dagger}\mathcal{D}_{n}=\Big(\frac{2\pi(n+\frac{1}{2})}{\beta}+\frac{2\pi w}{\beta}\Theta(x)\Big)^2-\partial_{x}^{2}+\frac{2\pi w}{\beta}\delta(x)\,.
\end{equation}
For our purpose, we want to compute the ratio
\begin{equation}
    \frac{\det(H_{n}+\lambda)}{\det(\frac{4\pi^2(n+\frac{1}{2})^2}{\beta^2}-\partial_{x}^2+\lambda)}\,.
\end{equation}
These ratios can be computed by the Gelfand-Yaglom method \cite{Kirsten:2005di} as
\begin{equation}
    \frac{\det(H_{n}+\lambda)}{\det(\frac{4\pi^2(n+\frac{1}{2})^2}{\beta^2}-\partial_{x}^2+\lambda)}=\frac{y_{1}(\frac{L}{2})}{y_{2}(\frac{L}{2})}\,,
\end{equation}
where we have
\begin{equation}
    (H_{n}+\lambda)y_{1}(x)=0\,,\quad y_{1}(-\frac{L}{2})=0\,,\quad y'_{1}(-\frac{L}{2})=1\,,
\end{equation}
\begin{equation}
    (\frac{4\pi^2(n+\frac{1}{2})^2}{\beta^2}-\partial_{x}^2+\lambda)y_{2}(x)=0\,,\quad y_{2}(-\frac{L}{2})=0\,,\quad y'_{2}(-\frac{L}{2})=1\,.
\end{equation}
It is easy to solve the above equations and obtain
\begin{equation}
    \frac{\det(H_{n}+\lambda)}{\det(\frac{4\pi^2(n+\frac{1}{2})^2}{\beta^2}-\partial_{x}^2+\lambda)}=\frac{d_{L}(r_n)}{d_{L}(p_n)}\frac{p_n\coth\frac{p_nL}{2}+r_n\coth\frac{r_nL}{2}+\frac{2\pi w}{\beta}}{2p_n\coth\frac{p_nL}{2}}\,,
\end{equation}
where we defined
\begin{equation}
    p_{n}=\sqrt{\frac{4\pi^2(n+\frac{1}{2})^2}{\beta^2}+\lambda}\,,\quad r_{n}=\sqrt{\frac{4\pi^2(n+w+\frac{1}{2})^2}{\beta^2}+\lambda}\,,\quad d_{L}(p)=\frac{\sinh\frac{pL}{2}}{p}\,.
\end{equation}
Thus, in the $L\rightarrow\infty$ limit all the $\coth$ becomes one, so we have
\begin{equation}
    \frac{\det(H_{n}+\lambda)}{\det(\frac{4\pi^2(n+\frac{1}{2})^2}{\beta^2}-\partial_{x}^2+\lambda)}=\frac{d_{L}(r_n)}{d_{L}(p_n)}\frac{p_n+r_n+\frac{2\pi w}{\beta}}{2p_n}\,.
\end{equation}
Thus, we have
\begin{equation}
    \frac{\det(\mathcal{D}^{\dagger}_{w}\mathcal{D}_{w}+\lambda)}{\det(\mathcal{D}^{\dagger}_{0}\mathcal{D}_{0}+\lambda)}=\prod_{n=-K}^{K}\frac{\det(H_{n}+\lambda)}{\det(\frac{4\pi^2(n+\frac{1}{2})^2}{\beta^2}-\partial_{x}^2+\lambda)}\,,\label{eq:ratio}
\end{equation}
where $K$ is a UV cutoff. Let's focus on the cases $w<0$ which are the cases for our exemplified calculation of the anomalous fermion correlators in Sec.~\ref{sec:correlators}. We are interested in the $\lambda\rightarrow0$ limit, under which the result of Equ.~(\ref{eq:ratio}) is
\begin{equation}
    \lim_{\lambda\rightarrow0}\frac{\det(\mathcal{D}^{\dagger}_{w}\mathcal{D}_{w}+\lambda)}{\det(\mathcal{D}^{\dagger}_{0}\mathcal{D}_{0}+\lambda)}=\lambda^{-w}e^{\frac{w(w+1)\pi L}{\beta}}\frac{\mathcal{T}_{-w}}{(\frac{4\pi}{\beta})^{-w}}\frac{\Gamma[K+\frac{1}{2}]}{\Gamma[K-w+\frac{1}{2}]}\,,
\end{equation}
where
\begin{equation}
    \mathcal{T}_{-w}=\big(\frac{-w}{\frac{4\pi}{\beta}}\big)^{-w}\Big(\frac{\Gamma[\frac{1}{2}]}{\Gamma[-w+\frac{1}{2}]}\Big)^2\,.
\end{equation}
Thus, when $K$ is large we have
\begin{equation}
    \lim_{\lambda\rightarrow0}\frac{\det(\mathcal{D}^{\dagger}_{w}\mathcal{D}_{w}+\lambda)}{\det(\mathcal{D}^{\dagger}_{0}\mathcal{D}_{0}+\lambda)}=\lambda^{-w}\mathcal{T}_{-w}\frac{e^{\frac{w(w+1)\pi L}{\beta}}}{(\frac{4\pi K}{\beta})^{-w}}\,.
\end{equation}
Since we want to remove the regulators $K$ and $L$ at the end, we have to deal with the factor 
\begin{equation}
 \frac{e^{\frac{w(w+1)\pi L}{\beta}}}{(\frac{4\pi K}{\beta})^{-w}}\,.   
\end{equation}
In fact, this factor can be made covariant if we use the heat kernel regularization. We will not go through the complete heat kernel regularization, but we will use it to covariantize the above factor by matching the heat kernel regularization to our cutoff regularization. This works as follows. Let's consider
\begin{equation}
    R(\lambda)\equiv\frac{\det(H_{w}+\lambda)}{\det(H_{0}+\lambda)}\,,\quad \text{where  }H_{w}=\mathcal{D}^{\dagger}_{w}\mathcal{D}_{w}\,\quad\text{and   } H_{0}=\mathcal{D}^{\dagger}_{0}\mathcal{D}_{0}\,.
\end{equation}
This ratio can be expressed using the heat kernel method as
\begin{equation}
    \log R(\lambda)=-\int_{0}^{\infty}\frac{dt}{t}e^{-\lambda t}\Tr(e^{-tH_{w}}-e^{-tH_{0}})\,.
\end{equation}
This expression is UV divergent because
\begin{equation}
    K_{w}(t)\equiv\Tr(e^{-tH_{w}}-e^{-tH_{0}})\rightarrow -\frac{w}{2}\,,\quad t\rightarrow0\,.
\end{equation}
Let's therefore regulate this divergence by an energy scale $\mu$ by considering
\begin{equation}
\begin{split}
\log R(\lambda,\mu)&=\log \frac{\det(\frac{H_{w}+\lambda}{\mu^2})}{\det(\frac{H_{0}+\lambda}{\mu^2})}\,,\\
&= -\frac{w}{2}\log\frac{\lambda}{\mu^2}-\int_{0}^{\infty}\frac{dt}{t}e^{-\lambda t} \big(K_{w}(t)+\frac{w}{2}\big)\,.\label{eq:regulatedR}
\end{split}
\end{equation}
Thus, we have
\begin{equation}
    R(\lambda)\rightarrow\big(\frac{\lambda}{\mu^2}\big)^{-\frac{w}{2}}\,,\quad\lambda\rightarrow\infty\,.\label{eq:Rlargelambda}
\end{equation}
Now we can match Equ.~(\ref{eq:Rlargelambda}) to the $\lambda\rightarrow\infty$ behavior of Equ.~(\ref{eq:ratio}). For finite $\lambda$, in the large $L$ and $K$ limit with $\frac{L}{\beta K^2}\ll1$ Equ.~(\ref{eq:ratio}) is\footnote{We note that this limit means that we take the limit $K\rightarrow\infty$ before we send the dimensionless box size $L/\beta$ to $\infty$ which is exactly the correct order limit for the above regularizations we did.}
\begin{equation}
\frac{\det(\mathcal{D}^{\dagger}_{w}\mathcal{D}_{w}+\lambda)}{\det(\mathcal{D}^{\dagger}_{0}\mathcal{D}_{0}+\lambda)}\rightarrow\frac{e^{\frac{w(w+1)\pi L}{\beta}}}{(\frac{4\pi K}{\beta})^{-w}}F_{w}(\lambda)\,.
\end{equation}
Then taking the large $\lambda$ limit we have\footnote{We note that this is the correct order of limit to match Equ.~(\ref{eq:Rlargelambda}) for which we also took $\lambda\rightarrow\infty$ at the end.}
\begin{equation}
\frac{\det(\mathcal{D}^{\dagger}_{w}\mathcal{D}_{w}+\lambda)}{\det(\mathcal{D}^{\dagger}_{0}\mathcal{D}_{0}+\lambda)}\rightarrow\frac{e^{\frac{w(w+1)\pi L}{\beta}}}{(\frac{4\pi K}{\beta})^{-w}}F_{w}(\lambda)\rightarrow \frac{e^{\frac{w(w+1)\pi L}{\beta}}}{(\frac{4\pi K}{\beta})^{-w}}\lambda^{-\frac{w}{2}}\,,   
\end{equation}
and matching to Equ.~(\ref{eq:Rlargelambda}) we have
\begin{equation}
    \mu=\frac{4\pi K}{\beta}e^{\frac{(w+1)\pi L}{\beta}}\,.
\end{equation}
Since we know that we have $-w$ zero modes of $\mathcal{D}_{w}$ and so $\mathcal{D}^{\dagger}_{w}\mathcal{D}_{w}$, the above covariant cutoff result implies
\begin{equation}
    \frac{\det'(\mathcal{D}^{\dagger}_{w}\mathcal{D}_{w})}{\det(\mathcal{D}^{\dagger}_{0}\mathcal{D}_{0})}=\lambda^{w}\lim_{\lambda\rightarrow0}\frac{\det(\mathcal{D}^{\dagger}_{w}\mathcal{D}_{w}+\lambda)}{\det(\mathcal{D}^{\dagger}_{0}\mathcal{D}_{0}+\lambda)}=e^{\frac{w(w+1)\pi L}{\beta}}\frac{\mathcal{T}_{-w}}{(\frac{4\pi K}{\beta})^{-w}}=\mu^{w}\mathcal{T}_{-w}\,.\label{eq:detratiosquare}
\end{equation}
Now let's compute the determinant of the zero modes Gram matrix $\det G^{f}[\alpha-\tilde{\alpha}]$. The zero mode functions in Equ.~(\ref{eq:zeromodes}) are greatly simplified if the periodic part of the rotor is removed. It is easy to show
\begin{equation}
    G_{rs}^{f}[\alpha-\tilde{\alpha}]=\frac{w\beta^2}{4\pi(r-\frac{w}{2})(r+\frac{w}{2})}\delta_{rs}\,,\label{eq:Gfinal}
\end{equation}
which implies
\begin{equation}
    \det G^{f}[\alpha-\tilde{\alpha}]=\beta^{-w}\mathcal{T}_{-w}\,.\label{eq:detG}
\end{equation}
Combining Equ.~(\ref{eq:detratiosquare}) and Equ.~(\ref{eq:detG}), we get
\begin{equation}
\frac{\det'(\mathcal{D}^{\dagger}_{w}\mathcal{D}_{w})}{\det(\mathcal{D}^{\dagger}_{0}\mathcal{D}_{0})\det G^{f}[\alpha-\tilde{\alpha}]}=\beta^{w}\mu^{w}\,.\label{eq:rationfinal}
\end{equation}
Thus, taking into account all of the flavors we have
\begin{equation}
    |\mathcal{N}(\beta)|^2=\Bigg |\frac{e^{\sum_{n>0}\frac{n}{2}\alpha_{n}\alpha_{-n}}\det'\mathcal{D}_{\alpha}}{\det \mathcal{D}_{0}\sqrt{\det G^{f}[\alpha]}}\Bigg|^{2N_{f}}=\beta^{N_{f}w}\mu^{N_{f}w}\,.
\end{equation}
Thus, we have
\begin{equation}
    \mathcal{N}(\beta)=e^{i\varphi}\beta^{\frac{N_{f}w}{2}}\mu^{\frac{N_{f}w}{2}}\,,
\end{equation}
where $\varphi$ is an unknown phase. Moreover, we should mention that the scale $\mu$ that sets the fermions' dimension in Sec.~(\ref{sec:anomalouscorrelators}) is actually the same as the $\mu$ here. This is because the regulated expression Equ.~(\ref{eq:regulatedR}) should be what actually appears in our calculation due to the Gaussian normalization of the path integral measure of all fermion modes with the scale $\mu$ as $\mu^2$ and the rescale of the Dirac operator by $\mu$.\footnote{Thus, we have
$\frac{\det' \mu D}{\det'\mu^2G_{\text{Gaussian}}}=\frac{\det' \frac{D}{\mu}}{\det'G_\text{Gaussian}}$ where $'$ means projecting to the space of all nonzero modes. With the Gaussian measure dropped in the ratio $\frac{\det'D}{\det' D_{0}}$, we therefore have $\frac{\det'\frac{D}{\mu}}{\det' \frac{D_{0}}{\mu}}$. Using the method similar to Equ.~(\ref{eq:detratiosquare}), we got $\frac{\det'\frac{D}{\mu}}{\det' \frac{D_{0}}{\mu}}=\mathcal{T}_{-w}$. Furthermore, Equ.~(\ref{eq:Gfinal}) would have an additional factor $\mu^2$. As a result, Equ.~(\ref{eq:ratio}) becomes $\frac{\det'(\frac{\mathcal{D}^{\dagger}_{w}\mathcal{D}_{w})}{\mu^2}}{\det(\frac{\mathcal{D}^{\dagger}_{0}\mathcal{D}_{0}}{\mu^2})\det G^{f}[\alpha-\tilde{\alpha}]}=\beta^{w}\mu^{w}$. As a result, this implies the result for $\mathcal{N}(\beta)$. } For notational convenience, we didn't retain it everywhere in the above discussions but instead introduced it as a regularization scale.

\bibliographystyle{JHEP}

\bibliography{main}

\end{document}